\documentclass[
reprint,
amsmath,amssymb,
aps,
pra,
]{revtex4-1}

\usepackage[dvipdfmx]{graphicx}
\usepackage{dcolumn}
\usepackage{bm}

\usepackage{color}
\usepackage{xr}

\begin{document}


\title{Spin-current instability at a magnetic domain wall in a ferromagnetic superfluid:\\
A generation mechanism of eccentric fractional skyrmions
 }

\author{Hiromitsu Takeuchi}
\email{takeuchi@osaka-cu.ac.jp}
\homepage{http://hiromitsu-takeuchi.appspot.com}
\affiliation{
Department of Physics and Nambu Yoichiro Institute of Theoretical and Experimental Physics (NITEP),\\
 Osaka City University, Osaka 558-8585, Japan
}




\date{\today}

\begin{abstract}
Spinful superfluids of ultracold atoms are ideal for investigating the intrinsic properties of spin current and texture because they are realized in an isolated, nondissipative system free from impurities, dislocations, and thermal fluctuations.
This study theoretically reveals the impact of spin current on a magnetic domain wall in spinful superfluids.
An exact wall solution is obtained in the ferromagnetic phase of a spin-1 Bose--Einstein condensate with easy-axis anisotropy at zero temperature.
The bosonic-quasiparticle mechanics analytically show that the spin current along the wall becomes unstable
if the velocity exceeds the critical spin-current velocities,
 leading to complicated situations because of the competition between transverse magnons and ripplons.
Our direct numerical simulation reveals that this system has a mechanism to generate an eccentric fractional skyrmion, which has a fractional topological charge, but its texture is not similar to that of a meron.
This mechanism is in contrast to the generation of conventional skyrmions in easy-axis magnets.
The theoretical findings can be examined in the same situation as in a recent experiment on ultracold atoms.
In terms of the universality of spontaneous symmetry breaking, unexplored similar phenomena are expected in different physical systems with the same broken symmetry.
\end{abstract}
\pacs{Valid PACS appear here}
\maketitle

\section{Introduction}
Membranal topological defects, called domain walls (DWs), are formed when a discrete symmetry is spontaneously broken in a phase transition.
They occur in every branch of physics, ranging from condensed matter fields, such as magnetism \cite{hubert2008magnetic} and optics \cite{gilles2017polarization}, to cosmology and field theory \cite{2006Vachaspati}.
DWs are also important in engineering because they appear in electronic and spintronic devices and magnetic materials, e.g., twisted nematic liquid crystals \cite{chandrasekhar_1992}, submicrometer ferromagnetic structures \cite{atkinson2003magnetic}, and ferromagnetic metals \cite{tatara1997resistivity}.

DWs in fermionic superfluids and superconductors with internal orbital and spin degrees of freedom have attracted increasing interest.
Chiral symmetry breaking has been confirmed in superfluid $^3$He-A  \cite{PhysRevLett.109.215301,ikegami2013chiral,ikegami2015observation},
and chiral domain structures have been observed in a slab geometry \cite{PhysRevLett.120.205301}.
Superfluid $^3$He-B is also capable of domain formation in a slab \cite{PhysRevLett.122.085301},
and an exotic DW that terminates on a half-quantum vortex as a nexus \cite{volovikPhysRevResearch.2.023263,Zhang2020PhysRevResearch.2.043356} has been realized in porous media \cite{makinen2019half}.
As fermionic quasiparticles form anomalous bound states at a DW or interface \cite{Aoki2005_PhysRevLett.95.075301,Nurakawa2009_PhysRevLett.103.155301,murakawa2010surface,Okuda2012_JPCM,Zheng2016_PhysRevLett.117.195301,Zheng2017_PhysRevLett.118.065301} due to a spatial modulation of the order parameter,
 the bound quasiparticles in superfluids and superconductors cause unconventional responses of currents \cite{matsumoto1999quasiparticle, Ashby2009PhysRevB.79.224509, Serban2010PhysRevLett.104.147001, Sauls2011PhysRevB.84.214509, Bouhon2014PhysRevB.90.220511,higashitani2014spin} and magnetic fields \cite{TanakaPhysRevB.72.140503,Yokoyama2011PhysRevLett.106.246601,Higashitani2013PhysRevLett.110.175301,Asano2015PhysRevB.92.224508}.

In contrast, DWs in bosonic superfluids and the properties associated with quasiparticle-bound states are less apparent.
Recently, composite defects of nematic-spin DWs and half-quantum vortices have been experimentally realized through the symmetry breaking phase transition in spin-1 Bose--Einstein condensates (BECs) of $^{27}$Na atoms \cite{kangPhysRevLett.122.095301}.
 The mass-current field around a composite defect is theoretically shown to form an anomalous elliptic structure in equilibrium \cite{TakeuchiPhysRevLett.126.195302}.
Magnetic-domain formation has been realized in a strongly ferromagnetic spinor BEC of $^{7}$Li atoms \cite{Huh2020PhysRevResearch.2.033471}.
 Spinor or multicomponent superfluids are suitable systems for exploring the intrinsic effect of spin currents because they are realized in isolated systems free from impurities and dislocations.
 A spin current (countersuperflow) has been experimentally shown to be intrinsically unstable without energy dissipation \cite{Hamner2011PhysRevLett.106.065302,KimPhysRevLett.119.185302} when the spin-current velocity exceeds the criteria \cite{Law2001PhysRevA.63.063612,Zhu2015PhysRevA.91.023633}.
 While these spin-current instabilities occur in the bulk without the involvement of DWs, the following fundamental questions have not been answered yet: How do spin currents influence a magnetic DW? What is the role of quasiparticle-bound states?

In this study, we theoretically investigate the properties of magnetic DWs with spin currents in the ferromagnetic phase of spin-1 BECs.
The DWs in this system are classified into antiferromagnetic (AF)-core and broken axisymmetry (BA)-core DWs according to the local magnetization at the wall (Fig.~\ref{Fig_DWprofile}).
The bosonic-quasiparticle mechanics reveal that a spin current causes spin-intrinsic instability in an AF-core DW,
 causing the condensation of the transverse magnon in a bound state to form a BA-core DW.
The instability dynamics have several channels to different results owing to the competition between the magnon condensation and Kelvin--Helmholtz instability (KHI),
which occur above the critical spin-current velocities.
Finally, our direct numerical simulation reveals that the instability at a BA-core DW causes eccentric fractional skyrmions, which are topological structures distinct from conventional skyrmions and merons.

This paper is organized as follows. We introduce the theoretical formalism basic to this study in Sec.~\ref{sec:basic}.
Section~\ref{sec:static} is devoted to the investigation of the static properties in the stationary state of a flat DW.
The excitation spectrum of a quasiparticle-bound state is investigated analytically in Sec.~\ref{sec:ES} and the critical spin-current velocity is computed in Sec.~\ref{sec:CspinV}.
In Sec.~\ref{sec:skyrmion}, it is numerically shown that eccentric fractional skyrmions are generated in quantum KHI of a BA-core DW.
A summary and prospects are made in the last section.

\section{Basic formulation}\label{sec:basic}
 Spin-1 BECs of dilute Bose gases at very low temperatures are described in the mean field theory, where the boson field operator $\hat{\Psi}_m^{(\dagger)}$ is replaced by the macroscopic wave function $\Psi_m^{(*)}$ in the many-body Hamiltonian. Here, $\Psi_m({\bm r},t)$ is the complex scalar field of the $\left|m\right>$ Zeeman ($m=0,\pm 1$) component.
  The dynamics in a uniform system are described by the Lagrangian functional of $\vec{\Psi}=[\Psi_{+1},\Psi_{0},\Psi_{-1}]^{\rm T}$ in the Gross--Pitaevskii (GP) model \cite{kawaguchi2012spinor} as
${\cal L}= \int d^3x  \left(\sum_{m}i\hbar\Psi_m^*\partial_t \Psi_m-{\cal G} \right)$,
with ${\cal G}( \vec{\Psi})=\frac{\hbar^{2}}{2M}\sum_{m}({\bm \nabla}\Psi_m^*)\cdot({\bm \nabla}\Psi_m)+ {\cal U}$ and
  \begin{eqnarray}
  {\cal U}=\frac{c_n}{2}n^2-\mu n
  +\frac{c_s}{2}{\bm s}^2-ps_z
  +\vec{\Psi}^{\dagger} (q\check{\sigma}_z^2+\check{U})\vec{\Psi}.
  \label{eq:energy_density}
  \end{eqnarray}
Here, we used the density
$n=\sum_m \left| \Psi_m \right|^2$
and the spin density
${\bm s}=[s_x,s_y,s_z]^{\rm T}= \sum_{mm'}\Psi_m^* \left(\check{\bm \sigma}\right)_{mm'} \Psi_{m'}$
 with the spin-1 matrices $\check{\bm \sigma}=[\check{\sigma}_x,\check{\sigma}_y,\check{\sigma}_z]^{\rm T}$,
 \begin{eqnarray}
   {\bm s}=\left(
   \begin{array}{c}
     s_x \\
     s_y \\
     s_z
   \end{array}
   \right)
   =\left(
   \begin{array}{c}
     \sqrt{2}{\rm Re}[(\Psi_{+1}+\Psi_{-1})\Psi_0^*] \\
     \sqrt{2}{\rm Im}[\Psi_0(\Psi_{+1}-\Psi_{-1})^*] \\
     |\Psi_{+1}|^2-|\Psi_{-1}|^2
   \end{array}
   \right)
   \label{eq:spin_density}
\end{eqnarray}
The Lagrange multipliers $\mu$ and $p$ are associated with the conservation of the particle number and longitudinal magnetization, respectively.
We neglect the external potential $\check{U}={\rm diag}(U_{+1},U_{0},U_{-1})$, unless otherwise noted.
Spin-1 BECs have four different phases: polar (P), AF, ferromagnetic (F), and BA phases \cite{kawaguchi2012spinor}.
Magnetic DWs are realized in the F phase with ferromagnetic interaction $-c_n < c_s<0$ and a negative quadratic Zeeman shift $q<0$.

The ground state $\vec{\Psi}=\vec{\Phi}_{\rm F\pm}$ in the F phase for $p\gtrless 0$ has a magnetization $s_z=\pm n$ and
 is represented as the F$_\pm$ state, $\Psi_{\pm 1}=\sqrt{\frac{\mu - q \pm p}{c_n+c_s}}e^{i\theta_{\rm G}}$ and $\Psi_{\mp 1}=\Psi_0=0$.
 Here, $\theta_{\rm G}=const.$ is the global phase according to the $U(1)$-symmetry breaking.
The energy densities ${\cal G}(\vec{\Phi}_{\rm F\pm})={\cal U}_{\rm F}=-\frac{1}{2}\frac{(\mu-q\pm p)^2}{c_n+c_s}$ of these states have the same value for $p=0$, corresponding to the spontaneous breaking of the discrete symmetry with respect to the spin inversion $s_z \leftrightarrow -s_z$.
 Magnetic DWs are topological defects associated with discrete symmetry breaking, in addition to quantized vortices with $U(1)$-symmetry breaking.
 The DW separates the domains in the two ground states with opposite magnetizations.
 The bulk ordered state is suppressed in the core of the DW,
  in which the macroscopic wave functions vary continuously between the two states.
The core of a topological defect in spin-1 BECs has been revealed to exhibit complicated states \cite{PhysRevA.86.013613, PhysRevLett.112.075301, PhysRevA.93.033633, weiss2019controlled, liu2020phase, underwood2020properties, PhysRevLett.125.170401, PhysRevLett.125.030402, PhysRevResearch.3.L012003, TakeuchiPhysRevLett.126.195302, 2021Takeuchi_PhysRevA.104.013316, Katsimiga_2021}.
Similarly, there are different possibilities for the local states at the wall in our system.


To approach the problem systematically, we introduce a general rule on the mass and spin currents along a DW.
This rule is a natural extension of the vortex winding rule for a rotational flow around an axisymmetric vortex \cite{isoshima2001quantum,TakeuchiPhysRevLett.126.195302,2021Takeuchi_PhysRevA.104.013316}.
The rule is applicable when all Zeeman components have finite population.

In the presence of currents along a flat DW normal to the $x$ axis,
the wave function in the stationary state is expressed as
$\Psi_m({\bm r},t)=f_m(x)e^{i \Theta_m}$
 with the real function $f_m$ and the phase
\begin{eqnarray}
 \Theta_m=\frac{\hbar}{M}{\bm V}_m\cdot {\bm r}+\vartheta_m,
\end{eqnarray}
where we used the current velocity ${\bm V}_m \bot \hat{\bm x}$ of the $m$ component defined by the current density $\mathbf{j}_{m}=\frac{\hbar}{M}{\rm Im}(\Psi_m^*{\bm \nabla}\Psi_m)=f_m^2{\bm V}_m$.
By substituting this formula into the equation of motion obtained from the Lagrangian,
we have the coupled equations
\begin{eqnarray}
&&\frac{\hbar^2}{2M}f_0''=g_0 f_{0} +c_{s} f_{0}\left( f^{2}_{+1} +f^{2}_{-1} +2f_{+1} f_{-1} e^{i\delta \Theta }\right)
\label{eq:f0_GP}\\
&&\frac{\hbar^2}{2M}f_{\pm 1}''=g_{\pm 1} f_{\pm 1} +c_{s} f^{2}_{0}\left( f_{\pm 1}+f_{\mp 1} e^{-i\delta \Theta } \right)
\label{eq:fpm_GP}
\end{eqnarray}
with $\delta \Theta=\Theta_{+1} + \Theta_{-1} - 2 \Theta_{0}$, $f_m''=\frac{{\rm d}^2}{{\rm d}x^2}f_m$, and
$$
g_m=c_{n} n +mc_{s} s_{z}-\mu-mp+m^2q +\frac{1}{2}M{\bm V}_m^2.
$$
When $f_m(x)$ is nonzero for all components by satisfying $f_{+1}f_{0}f_{-1}\neq 0$ at a certain place, $\delta \Theta$ must be an integer multiple of $\pi$ for satisfying the coupled Eqs.~(\ref{eq:f0_GP}) and (\ref{eq:fpm_GP}).
We may set as $\delta \Theta=0$ when the factor $e^{i\pi}=-1$ is included into the real function $f_m$ by changing its sign.
In this way, we have ${\bm V}_{+1} + {\bm V}_{-1}=2{\bm V}_0$ with $\vartheta_{+1}+\vartheta_{-1}-2\vartheta_{0}=0$, equivalent to the current velocity rule
\begin{eqnarray}
 {\bm V}_m={\bm V}_{+}+m{\bm V}_{-}.
 \label{eq:C_rule}
\end{eqnarray}
with ${\bm V}_{\pm}=\frac{1}{2}({\bm V}_{+1}\pm{\bm V}_{-1})$.
By substituting Eq.~(\ref{eq:C_rule}) into Eqs.~(\ref{eq:f0_GP}) and (\ref{eq:fpm_GP}) with $\delta\Theta=0$,
one obtains
\begin{eqnarray}
  0=H_mf_m+c_s(f_{+1}+f_{-1})^{2-m^2}f_0^{1+m^2}
  \label{eq:f_GPh}
\end{eqnarray}
with $H_m=-\frac{\hbar^2}{2M}\frac{{\rm d}^2}{{\rm d}x^2}+c_n n +mc_s s_z-\tilde{\mu}+m\tilde{p}+m^2\tilde{q}$
 and the hydrostatic variables \cite{2021Takeuchi_PhysRevA.104.013316},
\begin{eqnarray}
   &&\tilde{\mu}=\mu-\frac{1}{2}M{\bm V}_{+}^2
   \label{eq:hyst_mu}\\
   &&\tilde{p}=p- M {\bm V}_{+}\cdot{\bm V}_{-}
   \label{eq:hyst_p}\\
   &&\tilde{q}=q+\frac{1}{2}M{\bm V}_{-}^2
  \label{eq:hyst_q}
\end{eqnarray}
Equation~(\ref{eq:f_GPh}) is equivalent to Eqs.~(\ref{eq:f0_GP}) and (\ref{eq:fpm_GP}) with $\delta\Theta=0$ and ${\bm V}_m=0$ when $( \tilde{\mu}, \tilde{p},\tilde{q})$ are replaced by $(\mu, p, q)$;
the solution $f_m(x)$ of the former is equal to that of the latter when $( \tilde{\mu}, \tilde{p},\tilde{q})=(\mu, p, q)$.
Accordingly, we readjust the condition of the hydrostatic variables as $\tilde{\mu}>0$ and $\tilde{q}<0$ for realizing the F states in the bulk.

The bulk density is determined by the pressure balance between the two domains with opposite magnetization.
The hydrostatic pressure in a domain is computed in a similar manner as a scalar BEC \cite{pethick2008bose}.
 Since the hydrostatic pressure in a domain with $s_z=\pm n$ is given by $P_{\pm}=\frac{1}{2}\frac{\left( \tilde{\mu}-\tilde{q}\pm \tilde{p}\right)^2}{c_n+c_s}$,
the pressure equilibrium with a flat DW is realized under the condition $\tilde{p}=0$,
\begin{eqnarray}
P_+=P_-=P_{\rm F}\equiv \frac{1}{2}\frac{(\tilde{\mu}-\tilde{q})^2}{c_n+c_s}
\end{eqnarray}
with
\begin{eqnarray}
n(x\to \pm \infty) \to n_{\rm F}= \frac{\tilde{\mu}-\tilde{q}}{c_n+c_s},
\label{eq:nF}
\end{eqnarray}
or, equivalently, $s_z(x\to \pm \infty) \to \pm n_{\rm F}$.

\section{Static properties}\label{sec:static}
A DW solution in the F phase is obtained under the boundary condition of Eq.~(\ref{eq:nF}) for $-1<\frac{c_s}{c_n}< 0$ and $\tilde{q}<0$.
It is noted that a flat DW can be realized even for nonzero $p$ in the presence of currents by satisfying $\tilde{p}=0$.
The Galilean invariance allows us to set the center-of-mass velocity to be zero (${\bm V}_{+}=0$) without loss of generality in an isolated system.
In this work, hence, we consider the case with $p=0$ with ${\bm V}_{+}=0$,
 and then the DW solution is determined by the two parameters, $c_s/c_n$ and $q/\mu$ (or $\tilde{q}/\tilde{\mu}$) after rescaling the length, time, and wave function by
 $$\xi=\frac{\hbar}{\sqrt{M(\tilde{\mu}-\tilde{q})}},$$
 $$\tau=\frac{\hbar}{\tilde{\mu}-\tilde{q}},$$
  and $\sqrt{n_{\rm F}}$, respectively.

\begin{figure}
\begin{center}
\includegraphics[width=1.0 \linewidth, keepaspectratio]{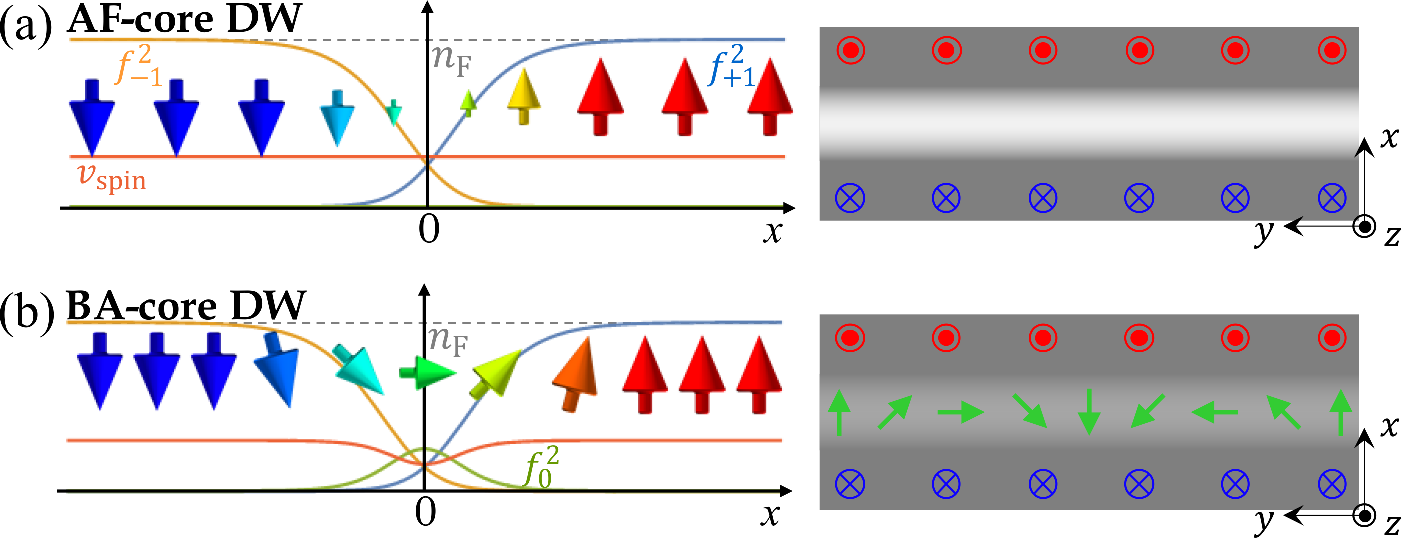}
\end{center}
\caption{
Schematic profiles of (a) an AF-core DW and (b) a BA-core DW. Left: The curves show the cross-sectional profile of the density $f_m^2~(m=0,\pm 1)$ and amplitude of the spin-current velocity $v_{\rm spin}$ parallel to the DWs for ${\bm V}_{+}=0$.
The arrows represent the spin texture.
Right: Two-dimensional profiles of spin density ${\bm s}$ (arrow) and density $n$ (gray scale).
}
\label{Fig_DWprofile}
\end{figure}

By changing $c_s/c_n$ and $\tilde{q}/\tilde{\mu}$ widely, we found two types of DW solutions, classified by the local ordered states at the center $x=0$:
 the AF-core DW with the local AF state [$f_{+ 1}(0)=f_{- 1}(0)$ and $f_0(0)=0$] and the BA-core DW with the local BA state [$f_{+ 1}(0)=f_{- 1}(0)$ and $f_0(0)\neq 0$] in Figs.~\ref{Fig_DWprofile}(a) and \ref{Fig_DWprofile}(b), respectively.
Here, we demonstrate the details of analytical and numerical results of DW solutions.

To discuss solutions of a AF-core DW,
 we first introduce the useful correspondence between binary BECs and spin-1 BECs in the absence of the $m=0$ component,
 which enable us to obtain some analytical expressions.
After the replacement of $(\Psi_{+1},\Psi_{-1})\to (\Psi_1,\Psi_2)$ and $(\mu +p-q,\mu -p-q)\to (\mu_1,\mu_2)$,
the Lagrangian of spin-1 BECs with $\Psi_0=0$ is rewritten as
${\cal L}= \sum_{j=1,2} \int d^3x  \left[ i\hbar\Psi_j^*\partial_t \Psi_j-\frac{\hbar^{2}}{2M}({\bm \nabla}\Psi_j^*)\cdot({\bm \nabla}\Psi_j)- {\cal U}_j \right]$ with
\begin{eqnarray}
{\cal U}_j= -\mu_j|\Psi_j|^2 + \sum_{k=1,2}\frac{g_{jk}}{2}|\Psi_j|^2|\Psi_k|^2 .
\label{eq:energy_density_binary}
\end{eqnarray}
This is just the Lagrangian of binary BECs with the intra- and inter-component coupling constants,
$g=g_{11}=g_{22}=c_n+c_s$ and $g_{12}=g_{21}=c_n-c_s$, respectively.
The ferromagnetic interaction $(0>c_s>-c_n)$ satisfies the immiscible condition of binary BECs, $0<g<g_{12}$,
 while the anti-ferromagnetic one $(c_s>0)$ does the miscible condition, $0<|g_{12}|<g$ \cite{pethick2008bose}.

 Joseph {\it et al.} pointed out that there is an exact solution of a DW in a segregated binary BEC with $g_{12}/g=3$ \cite{Indekeu2015_PhysRevA.91.033615}.
 This case corresponds to $c_n/c_s=-1/2$ in spin-1 BECs according to the above correspondence.
 Then the exact solution of the AF-core DW is given by
 \begin{eqnarray}
 f_{\pm 1}=f_{\pm 1}^{\rm ex}=\frac{\sqrt{n_{\rm F}}}{2}\left[1 \pm \tanh\left(\frac{x}{\xi} \right) \right].
 \label{eq:f1ex}
 \end{eqnarray}
 This solution is practical because the spin-1 BECs of $^{7}$Li atoms have $c_s \approx -c_n/2$ \cite{Huh2020PhysRevResearch.2.033471}.
 An exact DW solution with ferromagnetic interaction is also available for the BA phase \cite{Yu2021PhysRevResearch.3.023043,Yu2021anomalous}.

 \begin{figure}
 \begin{center}
 \includegraphics[width=1.0 \linewidth, keepaspectratio]{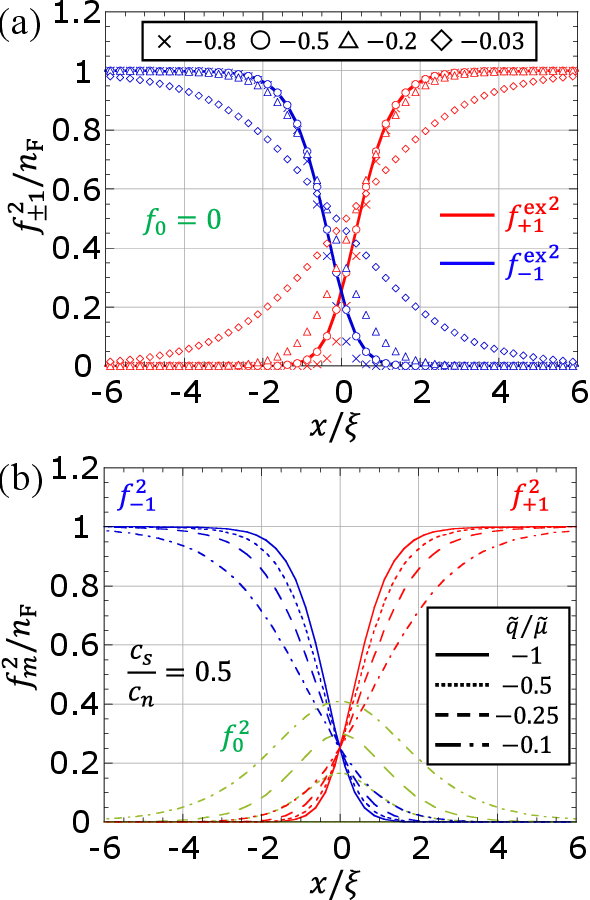}
 \end{center}
 \caption{
 (a) Numerical plots of density profiles $f_{+1}^2$ (red mark) and $f_{-1}^2$ (blue mark) of an AF-core DW for $c_s/c_n=-0.8,-0.5,-0.2,-0.03$.
 The dimensionless plots are independent of $\tilde{q}/\tilde{\mu}$.
 Solid curves show the exact solution $(f_{\pm 1}^{\rm ex})^2$ for $c_s/c_n=-0.5$ given by Eq.~(\ref{eq:f1ex}).
 (b) Numerical plots of $f_m^2~(m=0,\pm 1)$ of the lowest-energy DW for $\tilde{q}/\tilde{\mu}=-1,-0.5,-0.25,-0.1$ with $c_s/c_n=-0.5$.
 }
 \label{Fig_Nprofile}
 \end{figure}

Figure~\ref{Fig_Nprofile}(a) shows typical solutions of the AF-core DW obtained numerically by solving Eq.~(\ref{eq:nF}) for several values of $c_s/c_n$.
A stationary solution of a flat DW is obtained under the Neumann boundary condition at the $x=\pm L_x/2$ with the system size $L_x=256\xi$.
For details of the numerical method, see \ref{Asec:NMethod}.
The solutions of AF-core soliton depend on $\tilde{q}$ just through the rescaling parameters, $\xi$ and $n_F$, of the horizontal and vertical axes.
The $m=\pm 1$ components are likely to overlap and the density is almost homogeneous ($n\approx n_{\rm F}$) in the small-negative limit $c_s/c_n \to 0$,
 corresponding to the weak segregation limit with $g_{12}/g \to 1$ in binary BECs.
 The overlap is suppressed as $|c_s|/c_n$ increases.
The strong segregation limit $g_{12}/g \to \infty$ is realized for $c_s/c_n \to -1$,
 and then the overlap vanishes with $f_{+1}(0)f_{-1}(0)\to 0$.
In the following, we will focus mainly on the case of $c_s/c_n=-1/2$ and explain the numerical results only briefly for $c_s/c_n \neq -1/2$.

\begin{figure}
\begin{center}
\includegraphics[width=1.0 \linewidth, keepaspectratio]{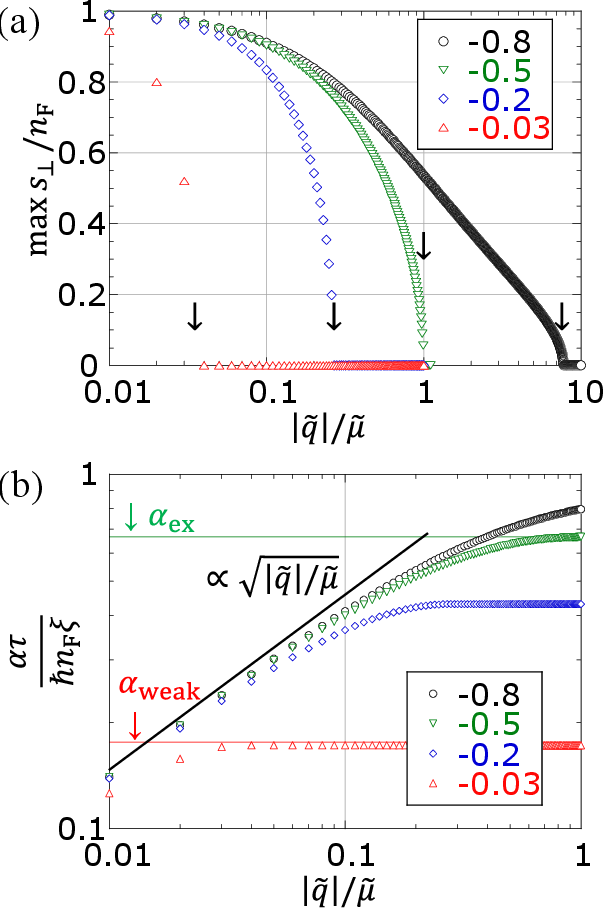}
\end{center}
\caption{
(a) The maximum transverse spin density $\max{s_\bot}$ and (b) the tension $\alpha$ of the lowest-energy DW for $c_s/c_n=-0.03,-0.2,-0.5,-0.8$.
(a) The lowest-energy DW is a BA-core DW with $\max{s_\bot}>0$ when $\tilde{q}$ is larger than a critical value $\tilde{q}_{\rm C}(c_s)<0$.
The positions of the transition points are indicated by arrows schematically.
(b) The analytical results, $\alpha_{\rm ex}$ [Eq.~(\ref{eq:a_ex})] and $\alpha_{\rm weak}$ [Eq.~(\ref{eq:a_weak})], for AF-core DWs are compared with the numerical results.
}
\label{Fig_maxSx}
\end{figure}

The transverse-spin density $s_\bot=\sqrt{s_x^2+s_y^2}$ is finite in the core of a BA-core DW with $f_0 \neq 0$, while ${\bm s}$ vanishes at the center $x=0$ of an AF-core DW with $f_0=0$ [see Eq.~(\ref{eq:spin_density})].
 Figure~\ref{Fig_Nprofile}(b) shows numerical solutions of a BA-core DW for several values of $\tilde{q}/\tilde{\mu}$ with $c_s/c_n$ fixed to be $-1/2$.
A continuous transition between BA-core and AF-core DWs occurs at $\tilde{q}=\tilde{q}_{\rm C}(c_s/c_n=-1/2)=-\tilde{\mu}$, which is predicted by the theoretical analysis demonstrated later.
Figure~\ref{Fig_maxSx}(a) demonstrates that the magnitude of $s_\bot$ is universally asymptotic to $n_{\rm F}$ in the limit of $\tilde{q}\to 0$ while the transition point, above which $s_\bot$ is nonzero, depends sensitively on $c_s/c_n$.
Therefore, the spin density ${\bm s}$ rotates but its amplitude is not so suppressed in a BA-core DW with small $|\tilde{q}|/\tilde{\mu}$.

The presence of the $m=0$ component in the core causes a difference in the profile of the spin-current velocity too (see left panels in Fig.~\ref{Fig_DWprofile}).
 The spin-current velocity ${\bm v}_{\rm spin}\equiv \frac{{\bm j}_{\rm spin}}{n}$, defined by the spin-current density ${\bm j}_{\rm spin}=\sum_m  m{\bm j}_m$ with ${\bm j}_m=\frac{\hbar}{M}{\rm Im}(\Psi_m^*{\bm \nabla}\Psi_m)$, is computed as
 \begin{eqnarray}
   {\bm v}_{\rm spin}=\frac{f_{+1}^2-f_{-1}^2}{n}{\bm V}_{+}+\frac{f_{+1}^2+f_{-1}^2}{n}{\bm V}_{-}.
 \end{eqnarray}
In our case of ${\bm V}_{+}=0$, the velocity reduces to ${\bm v}_{\rm spin}(x)=\frac{f_{+1}^2+f_{-1}^2}{n}{\bm V}_{-}$.
In an AF-core DW with $f_{+1}^2+f_{-1}^2=n$ with $f_0=0$,
the velocity is homogeneous, ${\bm v}_{\rm spin}={\bm V}_{-}=const.$.
In the BA-core DW, on the other hand, the spin-current velocity is locally suppressed in the core with $\frac{f_{+1}^2+f_{-1}^2}{n}<1$.
Since ${\bm s}$ rotates about the $z$ axis as $\Theta_{\pm 1}$ varies,
the spin texture takes a spiral structure on a BA-core DW in the presence of a finite spin current as shown schematically in the right panel of Fig.~\ref{Fig_DWprofile}(b).

The transition between BA-core and AF-core DWs occurs because the energy of the former is lower than that of the latter above the critical point: $\tilde{q}>\tilde{q}_{\rm C}$.
The energy of a DW is evaluated by the DW tension $\alpha$,
 which is important to determine its static and dynamic properties.
The tension $\alpha$ is defined by the excess energy in the presence of a DW of area $S$ as
\begin{eqnarray}
\alpha &=& S^{-1}\int d^3x\left[ {\cal G}-\tilde{\cal U}_{\rm b} \right]
\nonumber \\
       &=& \int_{-\infty}^{\infty} dx\left[ \frac{\hbar^2}{2M}\sum_mf_m'^2+U_{\rm pot} \right].
\end{eqnarray}
where $\tilde{\cal U}_{\rm b}=-P_{\rm F}$ is the energy density in the bulk.
The second line is obtained by substituting $\Psi_m=f_me^{i\Theta_m}$ into the first line and composed of the kinetic energy associated with the spatial gradient $f_m'=\frac{{\rm d}}{{\rm d}x}f_m$ and the residual term $U_{\rm pot}$.
Equation (\ref{eq:f_GPh}) is represented as $f_m'\frac{\hbar^2}{M}\frac{{\rm d}^2}{{\rm d}x^2}f_m'=f_m'\frac{\partial U_{\rm pot}}{\partial f_m}$.
By the integration and summation of this equation with respect to $x$ and $m$,
one obtains $\frac{\hbar^2}{2M}\sum_m f_m'^2=U_{\rm pot}$ with the boundary conditions $f_m'(\pm\infty)=0$ and $U_{\rm pot}(f_m(\pm \infty))=0$.
By using this result, the tension is represented by the formula
\begin{eqnarray}
\alpha =2\int_{-\infty}^{\infty} dx U_{\rm pot}= \frac{\hbar^2}{M}\sum_m \int_{-\infty}^{\infty} dx f_m'^2.
\end{eqnarray}
From the form on the rightmost side, the energy density at the wall is generally higher than that in the bulk and thus $n(x=0)$ is smaller than $n_{\rm F}$.

The tension $\alpha_{\rm AF}$ of an AF-core DW is evaluated analytically for some cases.
The tension $\alpha_{\rm ex}$ for the exact solution of Eq.~(\ref{eq:f1ex}) is computed as
\begin{eqnarray}
  \alpha_{\rm AF}=\alpha_{\rm ex}=\frac{2}{3}(\tilde{\mu}-\tilde{q})n_{\rm F}\xi~~~\left(\frac{c_s}{c_n}=-\frac{1}{2}\right).
  \label{eq:a_ex}
\end{eqnarray}
According to the correspondence with binary BECs,
we obtain the formulas of the tension in the weakly (strongly) segregating regime of $\frac{g_{12}}{g}-1 \ll 1$ $\left(\frac{g_{12}}{g}-1 \to \infty \right)$ \cite{Barankov2002_PhysRevA.66.013612,Schaeybroeck2008_PhysRevA.78.023624,Indekeu2015_PhysRevA.91.033615} as
\begin{eqnarray}
  &&\alpha_{\rm AF}=\alpha_{\rm weak}=\sqrt{\frac{-c_s}{c_n+c_s}}(\tilde{\mu}-\tilde{q})n_{\rm F}\xi~~~\left(\frac{|c_s|}{c_n}\ll 1\right),
  \label{eq:a_weak}\\
  &&\alpha_{\rm AF}= \frac{4}{3}(\tilde{\mu}-\tilde{q})n_{\rm F}\xi~~~\left(\frac{c_s}{c_n} \to -1\right).
  \label{eq:a_strong}
\end{eqnarray}

The tension $\alpha_{\rm BA}$ of a BA-core DW is computed by numerically minimizing $\alpha$, or equivalently, solving Eq.~(\ref{eq:f_GPh}).
Figure~\ref{Fig_maxSx}(b) shows the $\tilde{q}$-dependence of $\alpha_{\rm BA}$ for some typical values of $c_s$
 together with the plot of the peak value ($\max s_\bot$) of the transverse spin density at the DW.
For a fixed $c_s/c_n$, the tension $\alpha_{\rm BA}$ is smaller than $\alpha_{\rm AF}$ and approaches it for $\tilde{q} \to \tilde{q}_{\rm C}$,
at which a BA-core DW becomes an AF-core DW with $\max s_\bot=0$.
Therefore, the density $n$ in the core of a BA-core domain wall is less suppressed than a AF-core DW.
In the critical regime of $\tilde{q}/\tilde{\mu}\to 0$,
the tension is universally asymptotic to the scaling behavior $\frac{\alpha}{(\tilde{\mu}-\tilde{q})n_{\rm F}\xi}\propto \sqrt{|\tilde{q}|/\tilde{\mu}}$ independent of $c_s/c_n$ and then the density $n$ is almost constant, $n\approx n_{\rm F}$.

\section{Excitation spectrum}\label{sec:ES}
 The linear stability of a flat DW is evaluated by investigating excitation spectrum of bosonic quasiparticles.
 Here, we will demonstrate that excitations localized at the DW induces different types of spin-current instability.
 Depending on the properties of the localized excitations, called transverse magnons and ripplons, we introduce two different approaches.
 Especially, the analytical computations based on the two approaches give a quantitative prediction without fitting parameters for the ${^7}$Li case of $c_s/c_n=-1/2$ as shown in Fig.~\ref{Fig_BdG}.
 An approach for transverse magnons is a theoretical extension of the semiclassical approximation and the perturbation theory in quantum mechanics, called simply as the bosonic-quasiparticle mechanics in this paper.
  The other for ripplons is based on the low-energy effective theory that is a natural extension of the theoretical analysis of quantum KHI.
  We first discuss how a spin current influences the excitation spectrum for an AF-core DW based on the former approach.
  Then the latter one is introduced for describing the instability for AF-core and BA-core DWs in a unified manner.

\subsection{Bosonic-quasiparticle mechanics}
We obtain the excitation spectrum by solving the eigenvalue problem,
 derived by linearizing the equation of motion with respect to the Bogoliubov modes
$\delta\Psi_m({\bm r},t)=\Psi_m({\bm r},t)-\Phi_m({\bm r})=e^{i \Theta_m}\left[u_m(x)e^{i{\bm k}\cdot {\bm r}-i\omega t}-v_m(x)^*e^{-i{\bm k}\cdot {\bm r}+i\omega^* t}\right]$ with ${\bm k} \bot \hat{\bm x}$.
As the instability is induced by modes with nonpositive excitation energy,
we investigate the behavior of low-energy excitations.
For $f_0=0$, we have two independent eigenvalue equations for the vector fields $\vec{u}_0=(u_0,v_0)^{\rm T}$ and $(u_{+1},u_{-1},v_{+1},v_{-1})^{\rm T}$.
The problem with the latter is the same as quantum KHI in segregated binary BECs without external potentials \cite{Takeuchi2010PhysRevB.81.094517},
where the lowest-energy excitations of ripplons, the quanta of ripple waves on a DW, are described by the low-energy effective theory in Sec.~\ref{subsec:LET}.

The distinction from binary BECs is caused by the excitations of $\vec{u}_0$, called transverse magnons.
The magnons obey the eigenvalue equation
 \begin{eqnarray}
 \hbar \tilde{\omega} \vec{u}_0=
 \begin{pmatrix}
 h_0 +h_{\rm p} & h_{\rm p}\\
 -h_{\rm p} & -h_0 -h_{\rm p}
 \end{pmatrix}
 \vec{u}_0
 \label{eq:u_0}
 \end{eqnarray}
 with $\tilde{\omega}=\omega -{\bm k}\cdot {\bm V}_{+}$, $h_0=\frac{\hbar^2{\bm k}^2}{2M}+H_0+ c_s(f_{+1}+f_{-1})^2$, and $h_{\rm p}=-2c_sf_{+1}f_{-1}$.
 In our isolated system with Galilean invariance,
 we do not explicitly consider the center-of-mass motion causing the term $-{\bm k}\cdot {\bm V}_{+}$ on the right hand side of the equation of $\tilde{\omega}$.

Transverse magnons form bound states at the wall as follows.
In the classical limit of $-\frac{\hbar^2}{2M}\frac{{\rm d}^2}{{\rm d}x^2} \to \frac{P_x^2}{2M}$ in the semiclassical approximation of bosonic-quasiparticle mechanics \cite{WKB,takeuchi2018doubly},
we obtain the classical energy $E_{\rm classic}(P_x,x)$ as the eigenvalue for ${\bm k}=0$ as
\begin{eqnarray}
  E_{\rm classic}^2=\left(\frac{P_x^2}{2M}+V_+\right)\left( \frac{P_x^2}{2M}+V_-\right)
\end{eqnarray}
 with $V_{\pm}(x)=(c_n+c_s)n \pm 2c_sf_{+1}f_{-1}-\tilde{\mu}$ and the classical momentum $P_x$ in the $x$ direction.
 This approximation is a natural extension of the semiclassical approximation in quantum mechanics for a single-particle problem.
 In this analogy, the eigenvalue equation (\ref{eq:u_0}) corresponds to the ``Schr\"{o}dinger equation'' in quantum mechanics.
The classical energy can be lower than the energy gap $|\tilde{q}|$ of the magnon in the bulk,
\begin{eqnarray}
  E_{\rm classic}(0,x) =V_{\rm eff}(x)  \leq |\tilde{q}|,
\end{eqnarray}
 with
 \begin{eqnarray}
   V_{\rm eff}(x)\equiv \sqrt{\left[(c_s+c_n)n-\tilde{\mu}\right]^2-4c_s^2f_{+1}^2f_{-1}^2}.
\end{eqnarray}
 Here, we used the inequalities $n < n_{\rm F}$ and $f_{+1}f_{-1}\neq 0$ at the wall.
As the magnons must have an energy higher than $|\tilde{q}|$ to propagate in the bulk,
the bulk is the classically forbidden region for magnons with $E_{\rm classic}<|\tilde{q}|$,
 forming bound states at the wall.

The bound magnons can have even nonpositive or imaginary eigenvalues,
 indicating instability leading to its spontaneous excitation and the local condensation of the magnons.
 Therefore, the instability causes the transition of the DW to a BA-core DW by making the population of the $m=0$ component finite in the core.
The stability was evaluated using the criterion $V_{\rm eff}=0$ in the classical limit.
The semiclassical theory is applicable for the spin-1 BECs of $^{87}$Rb and $^{41}$K with a small negative $c_s/c_n$,
where the spatial variations of $n$ and $V_{\rm eff}$ are small, with $n\approx n_{\rm F}$.
Therefore, we may apply the approximation $V_{\rm eff}(0)^2 \approx \tilde{q}^2-\left( \frac{c_s}{c_n+c_s} \right)^2(\tilde{\mu}-\tilde{q})^2$ for $c_s\to 0$, and then the system becomes dynamically unstable with $E_{\rm classic}(0,x)$ imaginary if $\tilde{q} > \tilde{q}_{\rm C}$, where the critical value is given by
\begin{eqnarray}
  \tilde{q}_{\rm C} \approx \frac{c_s}{c_n}\tilde{\mu}~~~\left(\frac{|c_s|}{c_n}\ll 1\right).
  \label{eq:qc_scs}
\end{eqnarray}
 This result qualitatively explains the behavior of the transition point in the numerical plot of Fig.~\ref{Fig_maxSx}(a) for small $|c_s|/c_n$.
 This suggests that dynamic instability is induced by the quasiparticle-bound state under a spin current.
 The critical spin-current velocity will be defined in Sec.~\ref{sec:CspinV}.

To provide a quantitative understanding of this phenomenon beyond the semiclassical perspective, we focus on the excitations at the exact DW solution (\ref{eq:f1ex}) in $^{7}$Li condensates with $c_s=-\frac{c_n}{2}$.
The perturbation theory of bosonic-quasiparticle mechanics \cite{Skryabin2000PhysRevA.63.013602,Lundh2006PhysRevA.74.063620,takeuchi2018doubly,2021Takeuchi_PhysRevA.104.013316} is extended to calculate the criterion $\tilde{q}_{\rm C}(c_s=-\frac{c_n}{2})$.
This is a theoretical extension of the perturbation theory in quantum mechanics for a single-particle problem.

We consider the perturbation expansion in the following:
\begin{eqnarray}
  \hbar \tilde{\omega} \vec{u}_0=(\hat{h}_0+\delta\hat{h})\vec{u}_0
  \label{eq:BdGper}
\end{eqnarray}
with
$\displaystyle
 \hat{h}_{0} =\begin{pmatrix}
h_{0} & 0\\
0 & -h_{0}
\end{pmatrix}$
and
$\displaystyle
 \delta \hat{h} =\begin{pmatrix}
h_{\rm p} \  & h_{\rm p}\\
-h_{\rm p} & -h_{\rm p}
\end{pmatrix}$.
The unperturbed solution is given by the equations,
\begin{eqnarray}
  \begin{cases}
 \varepsilon u_{0} =h_{0} u_{0}\\
 \varepsilon v_{0} =-h_{0} v_{0}
 \end{cases}
 \label{eq:unper}
\end{eqnarray}
with the unperturbed eigenvalue $\varepsilon$.
These equations are solved exactly for the case of $c_s=-\frac{c_n}{2}$.
Then, substituting the exact DW solution of Eq.~(\ref{eq:f1ex}),
 Eqs.~(\ref{eq:unper}) are reduced to
\begin{eqnarray}
  \begin{cases}
  \epsilon_{+} u_{0} =\left[ -\frac{1}{2}\frac{d^{2}}{d(x/\xi)^{2}} -{\rm sech}^{2}\left(\frac{x}{\xi}\right) \right] u_{0}\\
  \epsilon_{-} v_{0} =\left[ -\frac{1}{2}\frac{d^{2}}{d(x/\xi)^{2}} -{\rm sech}^{2}\left(\frac{x}{\xi}\right) \right] v_{0}
  \end{cases}
  \label{eq:PTeq}
\end{eqnarray}
with $\epsilon_\pm=\frac{\tilde{\mu}-c_nn_{\rm F}/2-e_k\pm \varepsilon}{c_nn_{\rm F}/2}$ and $e_k=\frac{\hbar^2k^2}{2M}$ with $k=|{\bm k}|$.
The eigenvalue solution of Eq.~(\ref{eq:PTeq}) is given by solving the single-particle problem in the P\"{o}schl-Teller potential \cite{LandauQuantum}; $\epsilon_\pm=-\frac{1}{2}$ and $u_0, v_0 \propto {\rm sech}\left(\frac{x}{\xi}\right)$.
Accordingly, we have
\begin{eqnarray}
  \varepsilon=\hbar\omega_\pm=\pm(e_k-\tilde{\mu}+c_nn_{\rm F}/4).
\end{eqnarray}

\begin{figure*}
\begin{center}
\includegraphics[width=1.0 \linewidth, keepaspectratio]{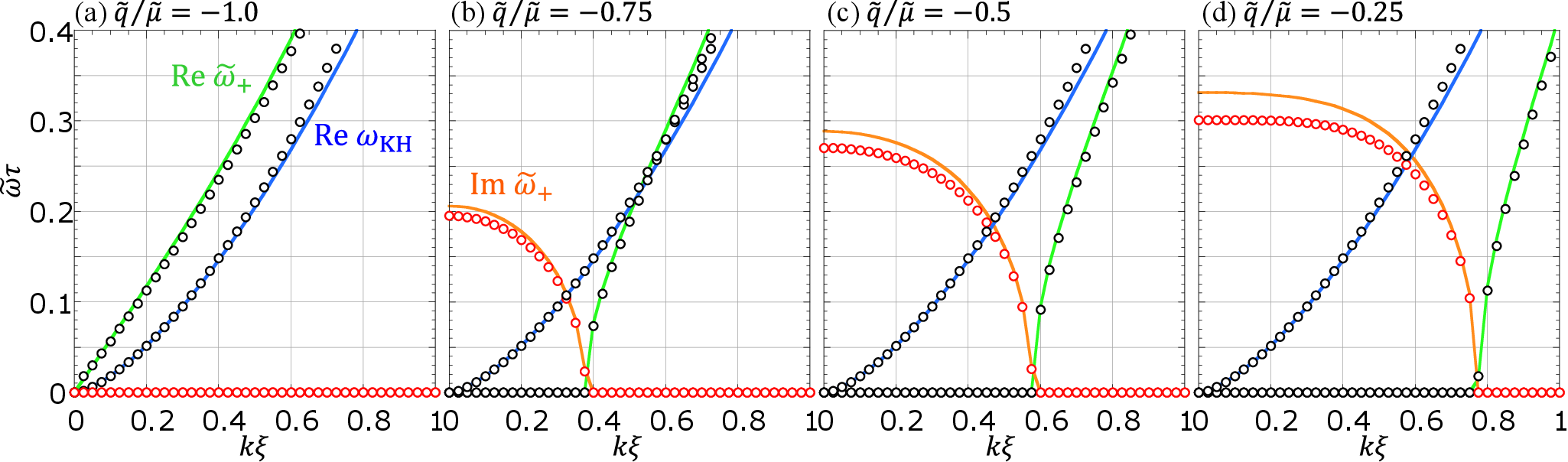}
\end{center}
\caption{
Bogoliubov excitation spectrum of the first and second lowest-energy excitations at an AF-core DW for $c_s/c_n=-1/2$ and ${\bm V}_+=0$. Circles show the real part (black circle) and the imaginary part (red circle) obtained numerically by solving the BdG equation.
Solid curves represent the analytical plots of ${\rm Re}~\tilde{\omega}_+$ (green) and ${\rm Im}~\tilde{\omega}_+$ (orange) in Eq.~(\ref{eq:w_per}) and ${\rm Re}~\omega_{\rm KH}$ (blue) in Eq.~(\ref{eq:WKH}) with $\alpha=\alpha_{\rm ex}$.
}
\label{Fig_BdG}
\end{figure*}

It is expected that the instability is induced by these bound states and the perturbed solution could be constructed by a combination of them in the two-mode approximation,
\begin{eqnarray}
\vec{u}_0=C_+\vec{u}_+ + C_-\vec{u}_-,
\label{eq:twoMode}
\end{eqnarray}
where we used the eigensolutions $(\varepsilon,\vec{u}_0)=(\hbar\omega_\pm,\vec{u}_\pm)$,
\begin{eqnarray}
  \begin{cases}
  \vec{u}_+ =\left[\frac{1}{\sqrt{2\xi}}{\rm sech}\left(\frac{x}{\xi}\right), 0\right]^{\rm T}\\
  \vec{u}_- =\left[0, \frac{1}{\sqrt{2\xi}}{\rm sech}\left(\frac{x}{\xi}\right) \right]^{\rm T}.
  \end{cases}
\end{eqnarray}
By inserting Eq.~(\ref{eq:twoMode}) into Eq.~(\ref{eq:BdGper}) and solving the resulting secular equation, one obtains
\begin{eqnarray}
\hbar\tilde{\omega}=\frac{\tilde{\varepsilon}_+ + \tilde{\varepsilon}_-}{2} \pm \sqrt{\left(\frac{\tilde{\varepsilon}_+ - \tilde{\varepsilon}_-}{2}\right)^2+M_{-+}M_{+-}}
\end{eqnarray}
with
$\tilde{\varepsilon}_\pm=\hbar\omega_\pm + M_{\pm\pm}$.
Here, $M_{\alpha\beta}$ is defined as
\begin{eqnarray}
  M_{\alpha \beta } =N_{\alpha \alpha }\int d^3x \vec{u}_{\alpha }^{\dagger } \hat{\sigma }_{z} \delta \hat{h} \vec{u}_{\beta}
\end{eqnarray}
with $N_{\alpha\beta}=\int d^3x \vec{u}_{\alpha }^{\dagger } \hat{\sigma }_{z}\vec{u}_{\beta}=\pm\delta_{\alpha\beta}$ and $\hat{\sigma }_{z}={\rm diag.}(1,-1)$.
The matrix elements $M_{\alpha\beta}$ were computed by using the normalization condition $N_{\pm \pm}=\pm 1$ as
\begin{eqnarray}
&&M_{++}=-M_{--}=M_{+-}=-M_{-+}=\frac{c_nn_{\rm F}}{6}.
\end{eqnarray}
Finally, we have $\tilde{\omega}=\pm{\omega}_{\rm mag}$:
\begin{eqnarray}
  \hbar \omega_{\rm mag}= \sqrt{ \left(\frac{\hbar^2k^2}{M}-\tilde{\mu}-\tilde{q}\right) \left( \frac{\hbar^2k^2}{4M}+\frac{\tilde{\mu}-7\tilde{q}}{12}\right)}.
\label{eq:w_per}
\end{eqnarray}
Figure~\ref{Fig_BdG} shows that the dispersion relation of Eq.~(\ref{eq:w_per}) quantitatively explains the results of the numerical diagonalization of the Bogoliubov--de Gennes (BdG) equation \cite{leggett2006quantum}, which is obtained after linearizing the equation of motion with respect to $u_m(x)$ and $v_m(x)$.
See \ref{Asec:NMethod} for details of the method of the numerical diagonalization.

Our theoretical analysis indicates that the system becomes dynamically unstable when the plus and minus branches of Eq.~(\ref{eq:w_per}) {\it collide} at $k=0$ by making a bubble of instability \cite{MacKay1987} when $\tilde{q}$ exceeds the critical value,
\begin{eqnarray}
  \tilde{q}_{\rm C}=-\tilde{\mu}~~~\left(\frac{c_s}{c_n}=-\frac{1}{2}\right).
  \label{eq:q_cLi}
\end{eqnarray}
The dispersion (\ref{eq:w_per}) is almost perfectly consistent with the numerical results near the critical point $\tilde{q}=-\tilde{\mu}$.
This consistency comes from the fact that the perturbed eigenvector $\vec{u}_0$ with $C_+=-C_-$ gives an exact solution for $\tilde{\omega}=k=0$ at the critical point at which we have $h_p=0$.

\subsection{Low-energy effective theory}\label{subsec:LET}
To investigate another dynamic instability induced by ripplons, we introduce the low-energy effective theory that describes the hydrodynamic aspect of a DW.
This theoretical framework is applicable to both the AF-core and BA-core DWs regardless of the internal structure of the core.
More specifically, the difference between the two is described through the tension $\alpha$ by neglecting the thickness of the DW.
A similar theory has been applied to a DW in binary BECs for describing quantum Kelvin-Helmholtz instability (KHI) \cite{Takeuchi2010PhysRevB.81.094517} and Nambu-Goldstone (NG) modes with the fractional dispersion \cite{TakeuchiPhysRevA.88.043612}.
Here, we extend the theory to spin-1 BECs.

The low-energy effective theory is constructed by considering the degrees of freedom associated with the NG modes and neglecting higher-energy Bogoliubov excitations.
We first consider excitations that propagate in the bulk far from the DW.
An excitation in the bulk is described as a collective fluctuation $\delta \vec{\Psi}=\vec{\Psi}-\vec{\Phi}_{\rm F}=\vec{u}-\vec{v}^*$ with the Bogoliubov coefficients $\vec{u}$ and $\vec{v}$.
The excitations can be classified by the coefficients, and
we have three types of excitations;
 e.g. in a magnetic domain with $s_z=+n$,
 (i) $\vec{u}\propto [1,0,0]^{\rm T}$ and $\vec{v}\propto [1,0,0]^{\rm T}$, (ii) $\vec{u}\propto [0,1,0]^{\rm T}$ and $\vec{v}=0$, and (iii) $\vec{u}\propto [0,0,1]^{\rm T}$ and $\vec{v}=0$.
Mode (i) is a gapless mode whose excitation spectrum $\varepsilon(k)$ does not have an energy gap for zero wave number ($k=0$): $\varepsilon(0)= 0$.
This mode is called phonon, corresponding to the NG mode associated with the $U(1)$ symmetry breaking.
The other two modes have energy gaps in the excitation spectrum: (ii) $\varepsilon(0)=|p|-q>0$ and (iii) $\varepsilon(0)=2|p|-2c_sn>0$.
Mode (ii) is equivalent to the transverse magnon, whose bound state was discussed in the bosonic-quasiparticle mechanics.
The Bogoliubov coefficients of (ii) and (iii) are composed of only the $m=0$ or $m=-1$ component,
 while those of (i) are composed of the $m=1$ component.
Therefore, we neglect the fluctuations of the $m=0$ and $m=-1$ components in the low-energy effective theory.
Similarly, in a domain with $s_z=-n$,
the degrees of freedom associated with the $m=0$ and $m=+1$ components are neglected approximately.

There exists another NG mode, called the ripplon, corresponding to the vibration mode of a DW.
According to the above consideration, we describe the state in a domain with $s_z=\pm n$ by a single wave function $\Psi_{\pm 1}$ in the effective theory.
Representing the position of the DW between two magnetic domains by the single-valued function $x=\eta(y,z,t)$,
we introduce the effective Lagrangian,
\begin{eqnarray}
  {\cal L}_{\rm eff}=\int {\rm d}y{\rm d}z\left( \int^{+\infty}_{\eta}{\cal P}_{+1}dx + \int^{\eta}_{-\infty}{\cal P}_{-1}dx \right)-\alpha S
  \nonumber
\end{eqnarray}
with the area $S$ and the tension $\alpha$ of the DW.
The Lagrangian density ${\cal P}_{\pm 1}$ in the domain with $s_z=\pm n$ is obtained by neglecting the contributions from the $m=0$ and $m=\mp 1$ components in the original Lagrangian ${\cal L}$;
$
{\cal P}_{\pm 1}=i\hbar\Psi_{\pm 1}^*\partial_t \Psi_{\pm 1}-\frac{\hbar^{2}}{2M}\sum_{m}({\bm \nabla}\Psi_{\pm 1}^*)\cdot({\bm \nabla}\Psi_{\pm 1})-{\cal U}_{\pm 1}
$
 with
$
{\cal U}_{\pm 1}=\frac{1}{2}(c_n+c_s)|\Psi_{\pm 1}|^4 -(\mu\pm p -q)|\Psi_{\pm 1}|^2.
$

By writing $S=\int {\rm d}x{\rm d}y \sqrt{1+(\partial_y \eta)^2+(\partial_z \eta)^2 } \approx \int {\rm d}x{\rm d}y \left[ 1+\frac{1}{2} (\partial_y\eta)^2+\frac{1}{2}(\partial_z\eta)^2 \right]$ for a small fluctuation from a stationary state of $\eta=0$,
one obtains the equation of motion with respect to $\eta$ from the effective Lagrangian,
\begin{eqnarray}
  {\cal P}_{+1}(\eta)-{\cal P}_{-1}(\eta) = \alpha \left(\partial_y^2+\partial_z^2 \right)\eta.
  \label{eq:YLeq}
\end{eqnarray}
This equation is an analog of the Young-Laplace equation in fluid mechanics, where the Laplace pressure ${\cal P}_{+1}-{\cal P}_{-1}$ balances the effect of the shape of the wall represented by the right-hand side of Eq.~(\ref{eq:YLeq}).
In the stationary state of $\eta=0$, the pressures in the two domains are equal with ${\cal P}_{+1}={\cal P}_{-1}=P_{\rm F}$.
The phonon is decoupled with the ripplon in the first approximation.
This is because the low-energy phonon can propagate through the wall without disturbance, associated with the anomalous tunneling effect of a DW \cite{PhysRevA.86.023622}.
Hence, the translational motion of a wall is treated independently from the bulk fluctuation and we mainly focus the former as a vibration mode of a DW.

The vibration of a DW is determined by Eq.~(\ref{eq:YLeq}) under a proper boundary condition on the wall.
The boundary condition can be different between our system and binary BECs,
while the theoretical treatment of the bulk wave function is identical to each other.
A possible mechanism to make the difference is the spin interaction.
The number of particles is conserved for each component in binary BECs, which demands the kinematic boundary condition \cite{2008Kundu},
whereas the population transfer occurs between different spin components via the spin interaction in spinor BECs.
However, we neglect the transfer in the effective theory for the following reason.
The mechanism of the transfer originally comes from the operator
 $\propto c_s \hat{\Psi}_{0}^{\dagger}\hat{\Psi}_{0}^{\dagger}\hat{\Psi}_{+1}\hat{\Psi}_{-1}$
  and its Hermitian conjugate in the spin interaction in the many-body Hamiltonian.
 The particle transfer can happen together with a pair annihilation or creation of the $m=0$ component, resulting in a fluctuation of the wave function $\Psi_0$.
 Such a process is forbidden at the DW in the low-energy effective theory because it can cost an extra energy by changing the internal structure of the wall, except for the bound magnon discussed above.
 Moreover, the ripplon is the NG mode associated with the spontaneous breaking of translational symmetry,
  and the fluctuation due to a ripple excitation makes a translational shift, expressed by $\delta\vec{\Psi}\propto \eta \frac{{\rm d}}{{\rm d}x}\vec{\Psi}$ by conserving the population of each component.
 We hence apply the same boundary condition as the one used in segregated binary BECs.

Once the correspondence between binary BECs and our system is established on the level of the effective theory,
 we can use the result of the dispersion relation of ripplons in binary BECs.
 Here, the roles of the first and second components in Refs.~\cite{Takeuchi2010PhysRevB.81.094517,TakeuchiPhysRevA.88.043612,tsubota2013quantum} are played by the $m=-1$ and $m=+1$ components in our system, respectively.
The dispersion $\omega_{\rm rip} ({\bm k})$ of a ripplon mode $\eta \propto \cos({\bm k}\cdot {\bm r}-\omega_{\rm rip} t)$ is given by
\begin{eqnarray}
\omega_{\rm rip}({\bm k},{\bm V}_{+},{\bm V}_{-})={\bm k}\cdot {\bm V}_{+} +\omega_{\rm KH}({\bm k},{\bm V}_{-})
\label{eq:Wrip}
\end{eqnarray}
with the KH dispersion
\begin{eqnarray}
\omega_{\rm KH}({\bm k},{\bm V}_{-})=\sqrt{\frac{\alpha |{\bm k}|^3}{2Mn_{\rm F}}-\left({\bm k}\cdot {\bm V}_{-} \right)^2}.
\label{eq:WKH}
\end{eqnarray}
The first term on the right hand side of Eq.~(\ref{eq:Wrip}) comes from the so-called Doppler effect induced by the center-of-mass motion related to the thermodynamic (or Landau) instability of the ripplon \cite{Takeuchi2010PhysRevB.81.094517}.

In the sense that the low-energy effective theory of KHI in binary BECs is identical to that for an AF-core DW in spinor BECs,
it can be said that the validity of the latter has been established partly by the numerical analyses in the previous works \cite{Takeuchi2010PhysRevB.81.094517,TakeuchiPhysRevA.88.043612}.
Here, for the purpose of reference, the prediction of the low-energy effective theory is evaluated just for ${\bm V}_+=0$ as plotted in Fig.~\ref{Fig_BdG}, where $\tilde{q}$ is changed with ${\bm V}_-$ fixed to be zero.
As $\frac{\alpha_{\rm AF}}{(\tilde{\mu}-\tilde{q})n_{\rm F}\xi}$ is independent of $\tilde{q}/\tilde{\mu}$ [see  Fig.~\ref{Fig_maxSx}(b) for $\tilde{q}>\tilde{q}_{\rm C}$],
the rescaled spectrum $\omega_{\rm KH}\tau$ with ${\bm V}_-=0$ is independent too.
The theory breaks down when the wave length becomes comparable to the thickness $\xi$ of an AF-core DW.
Accordingly, we see a little discrepancy between the numerical and theoretical plots for $k\xi \gtrsim 1$.

\section{Critical spin-current velocities}\label{sec:CspinV}
As phenomena described by the low-energy effective theory in our spinor BECs are the same as that in binary BECs,
no additional discoveries are expected even if we further proceed on the level of the linear stability analysis.
We find uniqueness rather when the core structure of the DWs undergoes a transition between AF-core and BA-core DWs in non-linear or nonequilibrium dynamics beyond the linear regime of the instability.
In this section, we illustrate that the competition between the transverse-magnon condensation and quantum KHI leads to an interesting scenario  of the instability development depending on the difference between their critical spin-current velocities.

\subsection{Criterion for transverse magnons}
The critical spin-current velocity for the transverse-magnon condensation is interpreted from the perspective of spin-current instability in the AF phase ($q<0$ and $c_s>0$) \cite{Zhu2015PhysRevA.91.023633, KimPhysRevLett.119.185302}.
The spin current in the AF state is dynamically unstable when the spin-current velocity is higher than \cite{Zhu2015PhysRevA.91.023633}
\begin{eqnarray}
  V_{\rm AF}(q)=\sqrt{\frac{2|q|}{M}},
  \label{eq:VAF}
\end{eqnarray}
 leading to the nucleation of the $m=0$ component owing to the collisional annihilation of the $m=\pm 1$ components.
A similar interpretation is applicable to the critical spin-current velocity $V_{\rm mag}$ for the local AF state at an AF-core DW.
By substituting $\tilde{q}=\tilde{q}_{\rm C}$ and $V_-=V_{\rm mag}$ into Eq.~(\ref{eq:hyst_q}),
we have
\begin{eqnarray}
V_{\rm mag}=\sqrt{\frac{2\tilde{q}_{\rm C}}{M}+V_{\rm AF}(q)^2}~~~(q \leq \tilde{q}_{\rm C}).
\label{eq:Vc}
\end{eqnarray}
An AF-core DW becomes dynamically unstable for $V_->V_{\rm mag}$ and then the transverse-magnon condensation leaves a BA-core DW by the phase transition from the local AF state to the BA state in the core with $s_\bot\neq 0$.
This is in contrast to the spin-current instability in the AF phase \cite{KimPhysRevLett.119.185302},
 which is interpreted as a phase transition from the AF phase ($\tilde{q}<0$) to the P phase ($\tilde{q}>0$) by ``quenching'' (rapidly increasing) the spin current velocity ($V_-$).

The critical spin-current velocity (\ref{eq:Vc}) is concretely computed for spinor condensates of $^{87}$Rb, $^{41}$K and $^7$Li atoms.
For $^{87}$Rb or $^{41}$K atoms, the result (\ref{eq:qc_scs}) by the semiclassical theory yields
\begin{eqnarray}
V_{\rm mag}= \sqrt{\frac{2}{M}\left(|q|+\frac{c_s}{c_n}\tilde{\mu}\right)}~~~\left( \frac{|c_s|}{c_n}\ll 1\right).
\label{eq:Vmag0}
\end{eqnarray}
This formula is consistently asymptotic to the criterion (\ref{eq:VAF}) by approaching the AF phase in the limit $c_s\to 0$.
By substituting Eq.~(\ref{eq:q_cLi}) into Eq.~(\ref{eq:Vc}), the criterion for the $^7$Li condensate is given by
 \begin{eqnarray}
   V_{\rm mag}=\sqrt{\frac{2}{M}(|q|-\tilde{\mu})}~~~\left(\frac{c_s}{c_n}=-\frac{1}{2}\right).
 \end{eqnarray}
 These results suggest that the critical point $\tilde{q}_{\rm C}$ and thus $V_{\rm mag}$ are monotonically increasing functions of $\frac{c_s}{c_n}$ for given $q$.
 This is because the transverse-magnon condensation is energetically preferred more for larger-negative $c_s$.

 \subsection{Criterion for ripplons}
 The critical spin-current velocity for quantum KHI is computed both for AF-core and BA-core DWs.
 Equation~(\ref{eq:WKH}) is regarded as the dispersion relation of capillary waves in fluid dynamics.
 Our system also supports an analog of the gravity-capillary waves more generally by introducing external potentials.
Such situations can be realized for a domain wall under a small magnetic-field gradient and a closed domain wall that surrounds a giant (multiply quantized) vortex \cite{HayashiPhysRevA.87.063628}.
Here, we consider the former case, where the potential for the $m$ component is represented as
$$U_{m}(x)=-m \frac{F}{2n_{\rm F}} x,$$
 which plays the role of a restoring force to stabilize the magnetic DW at the equilibrium position $\eta=0$.
According to the correspondence between binary and spinor BECs, the critical spin-current velocity for KHI is given by
\begin{eqnarray}
  V_{\rm rip}(\alpha) = \sqrt{\frac{\sqrt{F\alpha}}{Mn_{\rm F}}}.
\label{eq:V_rip}
\end{eqnarray}
A flat DW becomes dynamically unstable for $V_->V_{\rm rip}$.

The concrete formulas of $V_{\rm rip}$ are demonstrated for several cases as follows.
The critical spin-current velocity $V_{\rm rip}(\alpha_{\rm AF})$ for an AF-core DW is computed by using Eqs.~(\ref{eq:a_weak}) and (\ref{eq:a_ex}) as
\begin{eqnarray}
&&V_{\rm rip}(\alpha_{\rm weak}) = \frac{\xi}{\tau}\left(\sqrt{\frac{|c_s|}{c_n+c_s}}\frac{\tau\xi}{n_{\rm F}\hbar}F \right)^{\frac{1}{4}}~~~\left( \frac{|c_s|}{c_n}\ll 1\right)
\label{eq:V_ripAFweak}\\
&&V_{\rm rip}(\alpha_{\rm ex}) = \frac{\xi}{\tau}\left(\frac{2}{3}\frac{\tau\xi}{n_{\rm F}\hbar}F \right)^{\frac{1}{4}}~~~\left( \frac{|c_s|}{c_n}= -\frac{1}{2} \right)
\label{eq:V_ripAFex}
\end{eqnarray}
For computing the criterion for an-BA-core DW, we need to read the data of the tension $\alpha_{\rm BA}$ of a BA-core DW in Fig.~\ref{Fig_maxSx}(b).
It is convenient to use the critical behavior of the rescaled tension for $|\tilde{q}|/\tilde{\mu} \ll 1$,
\begin{eqnarray}
\frac{\alpha_{\rm BA} \tau}{\hbar n_{\rm F}\xi}\sim \sqrt{\frac{|\tilde{q}|}{\tilde{\mu}}}.
\label{eq:a_univ}
\end{eqnarray}
This scaling behavior is universally observed for different values of $c_s/c_n$.
As a result, one obtains the critical spin-current velocity for a BA-core DW,
\begin{eqnarray}
  V_{\rm rip}(\alpha_{\rm BA}) \sim \frac{\xi}{\tau}\left(\sqrt{\frac{|\tilde{q}|}{\tilde{\mu}}}\frac{\tau\xi}{n_{\rm F}\hbar}F \right)^{\frac{1}{4}}~~~\left( \frac{|\tilde{q}|}{\tilde{\mu}} \ll 1 \right).
\label{eq:V_ripBAuniv}
\end{eqnarray}

\subsection{Competition between two dynamic instabilities}
When only ripplons are excited at an AF-core DW in the spin-current instability and a BA-core DW is never formed,
the nonequilibrium development of the spin-current instability follows the regular scenario of quantum KHI;
according to Ref.~\cite{Takeuchi2010PhysRevB.81.094517}, quantized vortices are released from the DW in the nonequilibrium dynamics of KHI and the spin-current velocity (the relative velocity between superfluids in the two domains) decreases locally, leading to suppression of further instability.
However, the possibility of the BA-core formation due to the transverse-magnon condensation creates irregular scenarios.
The scenarios are generally complicated by the competition between the BA-core formation and KHI.

Let us consider the nonequilibrium development from the initial state of an flat AF-core DW.
As shown in Fig.~\ref{Fig_maxSx}(b),
the tension $\alpha_{\rm BA}$ of a BA-core DW, realized for $(\tilde{q}>\tilde{q}_{\rm C})$, is lower than that ($\alpha_{\rm AF}$) of an AF-core DW.
Therefore, if the BA-core formation occurs initially, the critical spin-current velocity becomes smaller than before: $V_{\rm rip}(\alpha_{\rm BA}) < V_{\rm rip}(\alpha_{\rm AF})$.
Then, KHI never occurs for $V_-<V_{\rm rip}(\alpha_{\rm BA})$ and the instability stops,
whereas it can occur again for $V_->V_{\rm rip}(\alpha_{\rm BA})$.
On the other hand, if ripplons rather than transverse magnons are excited initially with $V_->V_{\rm rip}(\alpha_{\rm AF})$,
quantized vortices are nucleated by following the regular scenario and the spin-current velocity is decreased to $V_-'(<V_-)$.
The BA-core formation can occur when $V_-'>V_{\rm mag}$ and the system follows the scenario mentioned first;
otherwise, for $V_-'<V_{\rm mag}$, the magnon condensation never occurs and the instability stops with $V_-'<V_{\rm rip}(\alpha_{\rm AF})$.



\section{Skyrmion generation}\label{sec:skyrmion}

Finally, we present an anomalous phenomenon as an incidental effect of the spin-current instability, found by the numerical experiments of quantum KHI at a BA-core DW.
We note that the spiraling spin texture along a BA-core DW [right panel in Fig.~\ref{Fig_DWprofile}(b)] is identical to the texture along a magnetic DW in the presence of a spin current in a magnet \cite{Kim2017PhysRevLett.119.047202},
where the spin current is metastable, leading to the generation of skyrmions with a unit topological charge.
The skyrmion generation from a magnetic DW in the magnetic system corresponds to the generation of skyrmions from a BA-core DW in our system.
However, interestingly, our spin-current instability generates not only skyrmions with a unit charge but also those with fractional charges.

\begin{figure*}
\begin{center}
\includegraphics[width=1.0 \linewidth, keepaspectratio]{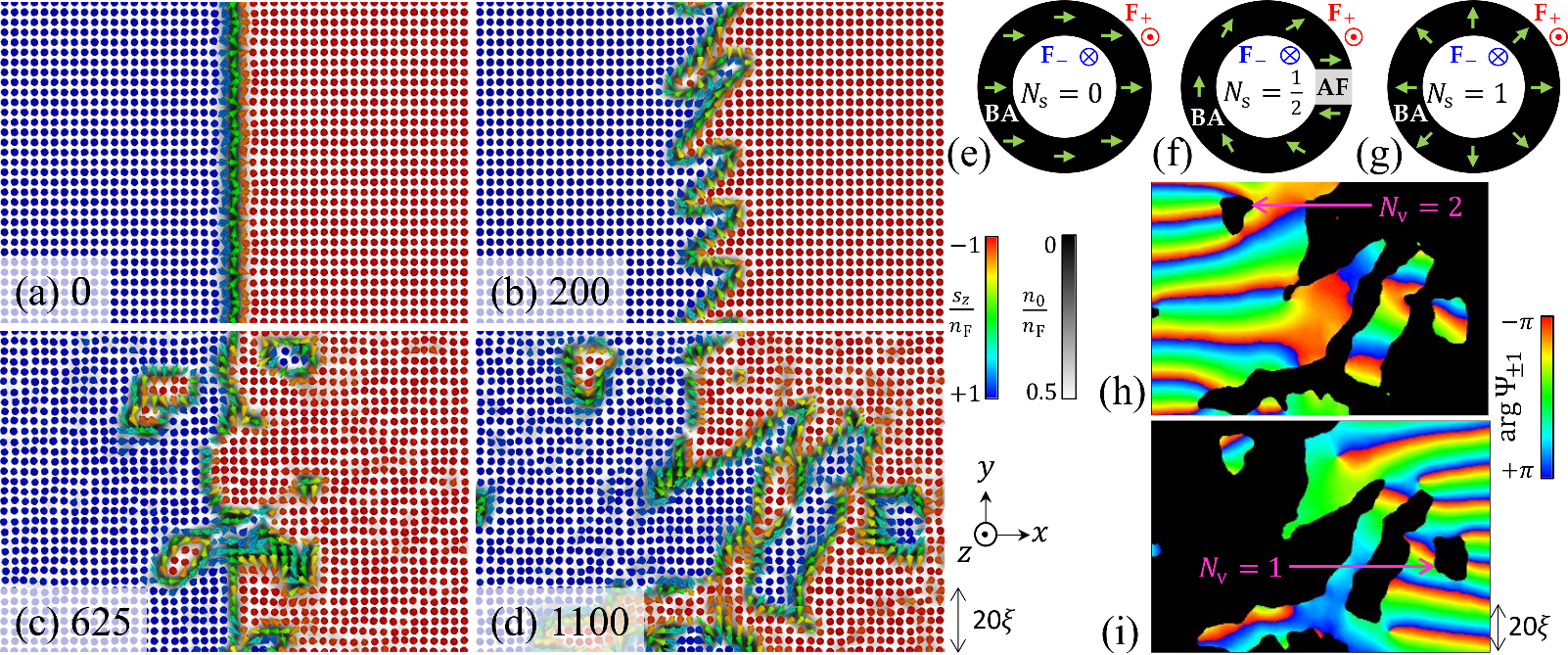}
\end{center}
\caption{
Dynamics of the spin-current instability at a BA-core DW for $(c_s/c_n, \tilde{q}/\tilde{\mu},V_{\pm1}\tau/\xi)=(-0.5,-0.05,\pm 0.368)$.
The arrows show the spin density ${\bm s}$.
The distributions of $s_z$ and $n_0=|\Psi_0|^2$ are represented by the arrow color and background grayscale, respectively.
(a) $t/\tau=0$: The wall is flat in the initial state.
(b) $t/\tau=200$: The KHI makes the flutter-finger pattern.
(c) $t/\tau=625$: The vortices are released as DW loops, and then spin singularities appear along the DWs at which spin density vanishes locally.
(d) $t/\tau=1100$: A spin singularity survives even on a released loop, forming a fractional skyrmion.
The spin texture of skyrmions is schematically illustrated for (e) $N_{\rm s}=0$, (f) $N_{\rm s}=1/2$, and (g) $N_{\rm s}=1$, together with the distribution of local ordered states (BA, AF, F$_\pm$).
The phases ${\rm arg}\Psi_{-1}$ and ${\rm arg}\Psi_{+1}$, of (d) are plotted in (h) and (i), respectively. The black regions represent $m=\pm 1$ domains, in which ${\rm arg}\Psi_{\mp 1}$ is highly fluctuated. The position of the domains with $N_{\rm v}=1,2$ is indicated by arrows.
}
\label{Fig_Cmark}
\end{figure*}

To demonstrate this anomaly effectively,
we numerically simulate quantum KHI at a flat BA-core domain wall in a uniform system with $U_m=0$ by solving the coupled GP equations for spin-1 BECs \cite{kawaguchi2012spinor}, the equation of motion obtained from the Lagrangian ${\cal L}$.
The initial state [Fig.~\ref{Fig_Cmark}(a)] of the time evolution is prepared by adding a small random fluctuation to the DW solution (see Appendix~\ref{Asec:NMethod} for details of the numerical method).
In the simplest case of quantum KHI with a small Weber number \cite{Kokubo2021_PhysRevA.104.023312},
the flutter-finger pattern of DW waves appears in the early stage [Fig.~\ref{Fig_Cmark}(b)].

An important distinction becomes apparent after the vortices are nucleated from the fingertips of DW waves.
 The vortex nucleation causes numerous spin singularities along the wall
 at which BA-core DWs are locally broken and replaced by AF-core DWs (the local AF state with ${\bm s}=0$ and $\Psi_0=0$) [Fig.~\ref{Fig_Cmark}(c)].
 Surprisingly, such singularities exist for a long time and survive even on the DW loops released into the bulk [see e.g., a DW loop on the far-right side of Fig.~\ref{Fig_Cmark}(d)].
 The time evolution of the whole system is demonstrated in Appendix~\ref{Asec:NMethod} including the explanations on the attached movie files of the animation in the Supplemental Material \cite{SM1}.
 It can be said that this phenomenon, caused by the density modulation of the $m=0$ component in the DW core, is unique to this system and cannot occur in quantum KHI in binary BECs.

 To explain the topological peculiarity of this phenomenon compared with the magnetic system \cite{Kim2017PhysRevLett.119.047202},
  we consider an isolated $m=\mp 1$ domain immersed in a sea of $m=\pm 1$ domains.
  The topology of the domains is classified by computing the Mermin--Ho relation \cite{Mermin1976_PhysRevLett.36.832.2}
  \begin{eqnarray}
    N_{\rm s}=\frac{1}{2}N_{\rm v}
  \label{eq:Skyrm}
  \end{eqnarray}
   between the skyrmion charge \cite{Skyrme1961_Proc.R.Soc.A}
   $$
   N_{\rm s}=\frac{1}{4\pi}\int_{\cal S} dxdy {\bm s}\cdot(\partial_x{\bm s} \times \partial_y {\bm s})
   $$  and the vortex winding number \cite{kawaguchi2012spinor}
   $$
   N_{\rm v}=\frac{M}{2\pi \hbar}\int_{\cal S} dxdy ({\bm \nabla}\times {\bm v}_{\rm mass})_z=\frac{M}{2\pi \hbar}\oint_{\cal C} d{\bm r}\cdot {\bm v}_{\rm mass}.
   $$
  Here, ${\cal S}$ is the surface enclosed by a closed contour ${\cal C}$ that surrounds the isolated domain, and
${\bm v}_{\rm mass}=\frac{\hbar}{Mn}\sum_m {\rm Im}(\Psi_m^*{\bm \nabla}\Psi_m)$ is the mass-current velocity.

 Although conventional skyrmions have an integer $N_{\rm s}$ under the boundary condition ${\bm s}(|{\bm r}|\to \infty)\parallel \pm \hat{\bm z}$,
 we may recognize the existence of a fractional skyrmion with $N_{\rm s}=1/2$ corresponding to the unit vortex charge $N_{\rm v}=1$.
 We typically observed isolated domains with $N_{\rm v}=0,1,2$ in the numerical simulations, and the
 corresponding skyrmions with $N_{\rm s}=0,1/2,1$ are illustrated in Figs.~\ref{Fig_Cmark}(e)--\ref{Fig_Cmark}(g).
 Domains of the integer skyrmion ($N_{\rm v}=2$) and fractional one ($N_{\rm v}=1$) displayed in Fig.~\ref{Fig_Cmark}(d) are indicated in the phase plots of Figs.~\ref{Fig_Cmark}(h) and \ref{Fig_Cmark}(i).
 For example, the solid angle covered by the spin texture in the $m=+1$ domain at the lower-left side of Fig.~\ref{Fig_Cmark}(c) is $4\pi$ (i.e., $N_{\rm s}=1$),
  which can be understood from the fact that ${\bm s}$ {\it winds} the equator {\it once} along the DW loop.
 In contrast, for an isolated domain with $N_{\rm s}=\frac{1}{2}$ on the far-right side of Fig.~\ref{Fig_Cmark}(d), ${\bm s}$ {\it winds} {\it half} of the equator along
the BA-core DW. Such a spin texture is realized in the presence of a spin singularity at the local AF state [see, also, Fig.~\ref{Fig_Cmark}(f)], where an energy divergence due to the spin singularity is avoided locally by removing the spin density with $n_0=|\Psi_0|^2=0$ while the density $n$ stays finite.
Since BA-core DWs are destroyed locally with spin singularity in a fractional skyrmion [Fig.~\ref{Fig_Cmark}(f)],
a fractional skyrmion will be observed as an image of ``C'' in the atomic cloud of the $m=0$ component or the transverse-spin distribution,
which is distinct from a conventional skyrmion that forms ``O'' [Figs.~\ref{Fig_Cmark}(e) and \ref{Fig_Cmark}(g)] (see, also, {\bf Movie S4} in the Supplemental Material \cite{SM1}).

Fractional skyrmions with spin singularity are in contrast to the meron or Mermin--Ho texture \cite{Parts1995_PhysRevLett.75.3320,volovik2003universe,girvin1999quantum,Moon1995_PhysRevB.51.5138,Kasamatsu2004_PhysRevLett.93.250406,kasamatsu2005vortices,Ezawa2011_PhysRevB.83.100408,Kharkov2017_PhysRevLett.119.207201,gao2019creation}.
 The former exists as an isolated object, whereas fractional charges in the latter are embedded in a periodic texture or realized only in pairs under the same boundary condition, e.g., recent observations in antiferromagnets \cite{gao2020fractional,jani2021antiferromagnetic}.
 Fractional skyrmions may be assumed to be unstable because spin singularity increases the energy cost.
 However, the cost is suppressed by the local AF (nematic-spin) order at the singularity.
 Furthermore, the splitting of an integer skyrmion with $N_{\rm s}=1$ into two isolated fractional skyrmions with $N_{\rm s}=1/2$ is preferred with respect to kinetic energy since two isolated vortices with $N_{\rm v}=1$ have lower kinetic energy than a vortex with $N_{\rm v}=2$ in the first approximation \cite{pethick2008bose}.
 Energetics are of fundamental importance because the relationship between the skyrmion charge and the vortex state with off-centered or eccentric spin singularity in Fig.~\ref{Fig_Cmark}(f) reminds us of that between the vortex winding rule and the nonaxisymmetic vortex \cite{TakeuchiPhysRevLett.126.195302,2021Takeuchi_PhysRevA.104.013316},
  which is related to the stability of the eccentric skyrmions.
  It is noted that similar objects have been discussed in superfluid $^3$He-A;
   ```SV'' (singular vortex) in the phase diagram of vortices \cite{Parts1995_PhysRevLett.75.3320,lounasmaa1999vortices}
    (see, also, \cite{seppala1983evidence,Seppala1984_PhysRevLett.52.1802,Volovic1984_PhysRevB.29.6090,Simola1987_PhysRevLett.58.904,fetter1987vortex}).
  Further discussion is beyond the scope of this work.

\section{Summary and Prospects}
In summary, we found two types of stationary solutions of a flat magnetic DW, called AF-core and BA-core DWs, in the F phase of spin-1 BECs.
The bosonic-quasiparticle mechanics based on the Bogoliubov theory revealed that the bound states of transverse magnons cause the spin-current instability at an AF-core DW, leading to the formation of a BA-core DW above the critical spin-current velocity for the magnons.
 We also extended the low-energy effective theory of ripplons to spinor BECs and predicted the critical spin-current velocity of quantum KHI for the ripplons.
The existence of two criteria for the spin-current velocity makes complex scenarios of the nonequilibrium development in the spin-current instability.
We numerically found that quantum KHI of a BA-core DW generates skyrmions of fractional topological charges with off-centered spin singularity, called eccentric fractional skyrmions.
Thanks to the universal applicability of the topological classification in terms of spontaneous symmetry breaking,
a similar but unexplored phenomenon is expected in different systems with the same broken symmetry.

Our theoretical predictions are expected to be examined with current experimental techniques.
As demonstrated in the experiment of spin-current instability in the AF phase \cite{KimPhysRevLett.119.185302},
a magnetic-field gradient induces a spin current along a magnetic DW in the F phase,
 which causes the transverse-magnon condensation and quantum KHI at the DW.
 Additionally, DWs are nucleated in the nonequilibrium process of spontaneous symmetry breaking in domain-coarsening dynamics in the F phase \cite{Williamson2016PhysRevLett.116.025301,Williamson2016PhysRevA.94.023608,Shitara_2017,Bourges2017_PhysRevA.95.023616}
  and spin currents occur in a complicated manner there.
  The occurrence of KHI there is justified by the quantum anomaly of the dynamic scaling behavior induced by vortex sheets (DWs with spin current) \cite{takeuchi2016domain,Bourges2017_PhysRevA.95.023616,TakeuchiPhysRevA.97.013617}.
  Such a domain coarsening dynamics is feasible if the ferromagnetic condensates are quenched into the deep F phase in the experiment \cite{Huh2020PhysRevResearch.2.033471}.
  If there exist only AF-core DWs in the early stage,
   the transverse-magnon condensation will be observed as the emergence of the $m=0$ component along the DW network in the atomic cloud.



\begin{acknowledgments}
This study was supported by JSPS KAKENHI Grants No. JP18KK0391, No. JP20H01842, and No. 20H01843
 and in part by the OCU ``Think globally, act locally '' Research Grant for Young Scientists through the hometown donation fund of Osaka City.
\end{acknowledgments}

\def\thesection{Appendix \Alph{section}}
\setcounter{section}{0}

\section{Numerical methods}\label{Asec:NMethod}

Here, we explain the method of numerical simulations used in this work.
All simulations are done by rescaling the length, time, and wave function by $\xi$, $\tau$, and $\sqrt{n_{\rm F}}$, respectively.

First, we describe how to obtain the DW solutions in Fig.~\ref{Fig_Nprofile}.
The stationary solution of a magnetic DW is obtained by solving Eq.~(\ref{eq:f_GPh})
 with the steepest descent method under the Neumann boundary condition
  $\left. \frac{ {\rm d} f_m }{ {\rm d}x}\right|_{x=\pm L_x/2}=0$ at the system boundary $x=\pm \frac{L_x}{2}$.
The space coordinate is discretized as $x\to x_i=-\frac{L_x}{2}+\Delta (i-1/2)~(i=0,1,2,...,N_x+1)$ with $N_x=1024$ and $\Delta=0.25 \xi$.
The spatial derivative of $f_m$ is computed with the finite difference approximation;
 $\frac{{\rm d}^2f_m}{{\rm d}x^2}$ is computed by the central difference of the second order.
The solutions of AF-core DWs are obtained by imposing the condition $f_0=0$.

The excitation spectrum in Fig.~\ref{Fig_BdG} is obtained by numerically diagonalizing the full BdG equations with respect to the eigenvector $[u_{+1}(x),u_{0}(x),u_{-1}(x),v_{+1}(x),v_{0}(x),v_{-1}(x)]^{\rm T}$ around the DW solution obtained above.
The spatial discretization, the spatial derivative, and the boundary condition for the eigenvector field are done in a similar way as described above.
The numerical diagonalization is performed with the double precise by using the Intel$\textsuperscript{\textregistered}$ Fortran Compiler with the Linear Algebra PACKage (LAPACK).

The result of Fig.~\ref{Fig_Cmark} (Fig.~\ref{Fig_p_Sbot} and the movie files; see Supplemental Material \cite{SM1}) is obtained by solving the GP equation in two dimensions.
The space coordinate is discretized as $(x,y)\to (x_i,y_j)=(-\frac{L}{2}+\Delta (i-1/2),-\frac{L}{2}+\Delta (j-1/2))~(i,j=0,1,2,...,N+1)$ with $N=512$ and $\Delta=0.5 \xi$.
Here, we impose the periodic boundary condition in the $y$ direction as $\Psi_m(x_i,y_{N+1})=\Psi_m(x_i,y_1)$ and $\Psi_m(x_i,y_0)=\Psi_m(x_i,y_N)$.
The spatial derivative is done in a similar way as described above.
The initial state of the time evolution is prepared by the steepest descent method.
The time $t$ is discretized as $t=\Delta t n_t$ with $\Delta t=0.0025 \tau$.
A small random fluctuation is added to the initial state to seed the instability.
The time evolution is computed by utilizing the Crank-Nicolson method.

\section{Instability dynamics of a BA-core DW}\label{Asec:IDBD}
\begin{figure*}
\begin{center}
\includegraphics[width=1.0 \linewidth, keepaspectratio]{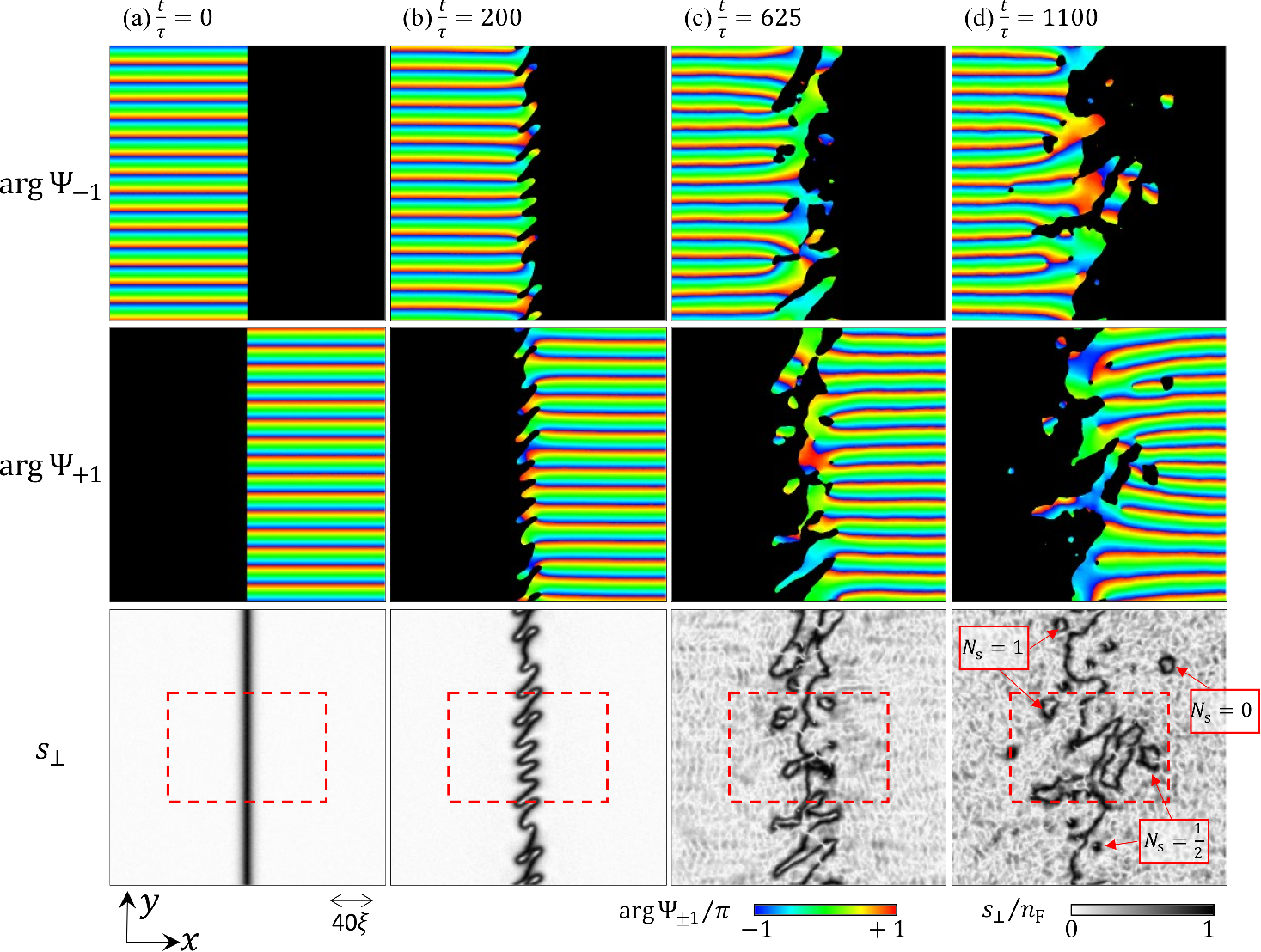}
\end{center}
\caption{
Snapshots of the phases $\arg\Psi_{\pm 1}$ and the transverse-spin density $s_\bot$ in the time evolution of Fig.~\ref{Fig_Cmark}.
The display range of Fig.~\ref{Fig_Cmark} is shown by red broken lines in the bottom panels.
Black areas in the top and middle panels represent the region of $s_z>0$ and $s_z<0$, respectively.
The transverse-spin density $s_\bot$ has a similar distribution to the background images of $n_0$ in Fig.~\ref{Fig_Cmark}.
In the later stage of the dynamics, a lot of ``notches'' appear as white spots along the DWs in the distribution of $s_\bot$, which correspond to the spin singularities in the spin density.
The number of branch cuts (jump from $\arg\Psi_{\pm 1}=-\pi$ to $\pi$) terminated by an isolated domain wall is equal to $N_{\rm v}(=N_{\rm s}/2)$.
A spin singularity exists along a closed DW surrounding an isolated domain, at which a branch cut is terminated.
The closed DW with a spin singularity looks like a ``C'' mark in the distribution plot of $s_\bot$ or $n_0$.
No singularity happen if the domain contains an even number of terminated branch cuts.
Some examples of domains with $N_{\rm s}=0,1/2,1$ are denoted by red arrows in the lower right.
}
\label{Fig_p_Sbot}
\end{figure*}

Here, we explain the detailed information on the numerical result of the spin-current instability demonstrated in Fig.~\ref{Fig_Cmark}.
The images in Fig.~\ref{Fig_Cmark}(a)--\ref{Fig_Cmark}(d) are magnified snapshots of a numerical simulation.
Figure \ref{Fig_p_Sbot} shows the time evolution of the phases $\arg\Psi_{\pm 1}$ and the transverse-spin amplitude $s_\bot$ in the same simulation.

The attached movie files (see Supplemental Material \cite{SM1}) demonstrate the animations of different quantities made with 51 snapshots from $t/\tau=0$ to $t/\tau=1250$ in the same simulation. The box size is the same as the system size of the simulation.
Figures \ref{Fig_Cmark}(a)--\ref{Fig_Cmark}(d) are magnified images of some snapshots in the animation of {\bf Movie S1}.
The top ($\arg\Psi_{-1}$), middle ($\arg\Psi_{+1}$), and bottom ($s_\bot$) panels in Fig.~\ref{Fig_p_Sbot} are some snapshots in the animations of {\bf Movie S2},  {\bf Movie S3}, and  {\bf Movie S4}, respectively.
The plot area of Fig.~\ref{Fig_Cmark}(a)--\ref{Fig_Cmark}(d) is implied by dashed squares in the bottom panels in Fig.~\ref{Fig_p_Sbot}.

Some isolated domains with different topological charges are implied by arrows in the bottom right in Fig.~\ref{Fig_p_Sbot}.
According to Eq.~(\ref{eq:Skyrm}), we can convert the number $N_{\rm v}$ of branch cuts (jump from $\arg\Psi_{\pm 1}=-\pi$ to $\pi$), which are terminated by an isolated domain, to the skyrmion charge $N_{\rm s}(=2N_{\rm v})$ of the domain.
We see that a branch cut passes through an isolated domain corresponding to the case of $N_{\rm s}=0$ from the comparison between the distributions of $\arg\Psi_{+1}$ and $s_\bot$ at $t/\tau=1100$.
Two branch cuts end at a domain by forming a conventional skyrmion with $N_{\rm s}=1$.
An isolated domain that terminates a branch cut corresponds to an eccentric fractional skyrmion with $N_{\rm s}=1/2$.
The spin singularities appear in the form of cut points of a BA-core DW or white spots in the distribution of $s_\bot$ in the bottom panels of Fig.~\ref{Fig_p_Sbot}.
According to this property, an eccentric fractional skyrmion will be observed as an image of ``C'' in the atomic cloud of the $m=0$ component or the distribution of $s_\bot$.
In contrast, skyrmions with integer or zero charges make images of ``O''.
This contrasting behavior between the novel and conventional skyrmion is easily observed in {\bf Movie S4} (see Supplemental Material \cite{SM1}).

\begin{thebibliography}{10}

\bibitem{hubert2008magnetic}
Alex Hubert and Rudolf Sch{\"a}fer.
\newblock {\em Magnetic domains: The Analysis of Magnetic Microstructures}.
\newblock Springer Science \& Business Media, New York, 2008.

\bibitem{gilles2017polarization}
Marin Gilles, P-Y Bony, Jocelin Garnier, Antonio Picozzi, Massimiliano Guasoni,
  and Julien Fatome.
\newblock {Polarization domain walls in optical fibres as topological bits for
  data transmission}.
\newblock {\em Nat. Photon.}, 11(2):102--107, 2017.

\bibitem{2006Vachaspati}
T.~Vachaspati.
\newblock {\em Kinks and Domain Walls: An Introduction to Classical and Quantum
  Solitons}.
\newblock Cambridge University Press,Cambridge, 2006.

\bibitem{chandrasekhar_1992}
S.~Chandrasekhar.
\newblock {\em Liquid Crystals}.
\newblock Cambridge University Press, 2 edition, 1992.

\bibitem{atkinson2003magnetic}
Del Atkinson, Dan~A Allwood, Gang Xiong, Michael~D Cooke, Colm~C Faulkner, and
  Russell~P Cowburn.
\newblock {Magnetic domain-wall dynamics in a submicrometre ferromagnetic
  structure}.
\newblock {\em Nat. Mater.}, 2(2):85--87, 2003.

\bibitem{tatara1997resistivity}
Gen Tatara and Hidetoshi Fukuyama.
\newblock {Resistivity due to a domain wall in ferromagnetic metal}.
\newblock {\em Physical Review Letters}, 78(19):3773, 1997.

\bibitem{PhysRevLett.109.215301}
P.~M. Walmsley and A.~I. Golov.
\newblock {Chirality of superfluid $^{3}\mathrm{He}$-$A$}.
\newblock {\em Phys. Rev. Lett.}, 109:215301, Nov 2012.

\bibitem{ikegami2013chiral}
H~Ikegami, Y~Tsutsumi, and K~Kono.
\newblock {Chiral symmetry breaking in superfluid 3He-A}.
\newblock {\em Science}, 341(6141):59--62, 2013.

\bibitem{ikegami2015observation}
Hiroki Ikegami, Yasumasa Tsutsumi, and Kimitoshi Kono.
\newblock {Observation of intrinsic magnus force and direct detection of
  chirality in superfluid 3He-A}.
\newblock {\em Journal of the Physical Society of Japan}, 84(4):044602, 2015.

\bibitem{PhysRevLett.120.205301}
J.~Kasai, Y.~Okamoto, K.~Nishioka, T.~Takagi, and Y.~Sasaki.
\newblock {Chiral Domain Structure in Superfluid
  $^{3}\mathrm{He}\text{\ensuremath{-}}A$ Studied by Magnetic Resonance
  Imaging}.
\newblock {\em Phys. Rev. Lett.}, 120:205301, May 2018.

\bibitem{PhysRevLett.122.085301}
Lev~V. Levitin, Ben Yager, Laura Sumner, Brian Cowan, Andrew~J. Casey, John
  Saunders, Nikolay Zhelev, Robert~G. Bennett, and Jeevak~M. Parpia.
\newblock {Evidence for a Spatially Modulated Superfluid Phase of
  $^{3}\mathrm{He}$ under Confinement}.
\newblock {\em Phys. Rev. Lett.}, 122:085301, Feb 2019.

\bibitem{volovikPhysRevResearch.2.023263}
G.~E. Volovik and K.~Zhang.
\newblock {String monopoles, string walls, vortex skyrmions, and nexus objects
  in the polar distorted B phase of $^{3}$He}.
\newblock {\em Phys. Rev. Res.}, 2:023263, Jun 2020.

\bibitem{Zhang2020PhysRevResearch.2.043356}
K.~Zhang.
\newblock {One-dimensional nexus objects, network of Kibble-Lazarides-Shafi
  string walls, and their spin dynamic response in polar-distorted $B$-phase of
  $^{3}\mathrm{He}$}.
\newblock {\em Phys. Rev. Res.}, 2:043356, Dec 2020.

\bibitem{makinen2019half}
JT~M{\"a}kinen, VV~Dmitriev, Jaakko Nissinen, Juho Rysti, GE~Volovik, AN~Yudin,
  Kuang Zhang, and VB~Eltsov.
\newblock {Half-quantum vortices and walls bounded by strings in the
  polar-distorted phases of topological superfluid $^3$He}.
\newblock {\em Nat. Commun.}, 10(1):237, 2019.

\bibitem{Aoki2005_PhysRevLett.95.075301}
Y.~Aoki, Y.~Wada, M.~Saitoh, R.~Nomura, Y.~Okuda, Y.~Nagato, M.~Yamamoto,
  S.~Higashitani, and K.~Nagai.
\newblock {Observation of Surface Andreev Bound States of Superfluid
  $^{3}\mathrm{He}$ by Transverse Acoustic Impedance Measurements}.
\newblock {\em Phys. Rev. Lett.}, 95:075301, Aug 2005.

\bibitem{Nurakawa2009_PhysRevLett.103.155301}
S.~Murakawa, Y.~Tamura, Y.~Wada, M.~Wasai, M.~Saitoh, Y.~Aoki, R.~Nomura,
  Y.~Okuda, Y.~Nagato, M.~Yamamoto, S.~Higashitani, and K.~Nagai.
\newblock {New Anomaly in the Transverse Acoustic Impedance of Superfluid
  $^{3}\mathrm{He}\mathrm{\text{\ensuremath{-}}}\mathbf{B}$ with a Wall Coated
  by Several Layers of $^{4}\mathrm{He}$}.
\newblock {\em Phys. Rev. Lett.}, 103:155301, Oct 2009.

\bibitem{murakawa2010surface}
Satoshi Murakawa, Yuichiro Wada, Yuta Tamura, Masahiro Wasai, Masamichi Saitoh,
  Yuki Aoki, Ryuji Nomura, Yuichi Okuda, Yasushi Nagato, Mikio Yamamoto, et~al.
\newblock {Surface Majorana cone of the superfluid 3He B phase}.
\newblock {\em Journal of the Physical Society of Japan}, 80(1):013602, 2010.

\bibitem{Okuda2012_JPCM}
Okuda Y and Nomura R.
\newblock {Surface Andreev bound states of superfluid3He and Majorana
  fermions}.
\newblock {\em Journal of Physics: Condensed Matter}, 24(34):343201, aug 2012.

\bibitem{Zheng2016_PhysRevLett.117.195301}
P.~Zheng, W.~G. Jiang, C.~S. Barquist, Y.~Lee, and H.~B. Chan.
\newblock {Anomalous Damping of a Microelectromechanical Oscillator in
  Superfluid $^{3}\mathrm{He}$-B}.
\newblock {\em Phys. Rev. Lett.}, 117:195301, Nov 2016.

\bibitem{Zheng2017_PhysRevLett.118.065301}
P.~Zheng, W.~G. Jiang, C.~S. Barquist, Y.~Lee, and H.~B. Chan.
\newblock {Critical Velocity in the Presence of Surface Bound States in
  Superfluid $^{3}\mathrm{He}\text{\ensuremath{-}}\mathrm{B}$}.
\newblock {\em Phys. Rev. Lett.}, 118:065301, Feb 2017.

\bibitem{matsumoto1999quasiparticle}
Masashige Matsumoto and Manfred Sigrist.
\newblock {Quasiparticle states near the surface and the domain wall in
  apx$\pm$ipy-wave superconductor}.
\newblock {\em Journal of the Physical Society of Japan}, 68(3):994--1007,
  1999.

\bibitem{Ashby2009PhysRevB.79.224509}
Phillip E.~C. Ashby and Catherine Kallin.
\newblock {Suppression of spontaneous supercurrents in a chiral $p$-wave
  superconductor}.
\newblock {\em Phys. Rev. B}, 79:224509, Jun 2009.

\bibitem{Serban2010PhysRevLett.104.147001}
I.~Serban, B.~B\'eri, A.~R. Akhmerov, and C.~W.~J. Beenakker.
\newblock Domain wall in a chiral $p$-wave superconductor: A pathway for
  electrical current.
\newblock {\em Phys. Rev. Lett.}, 104:147001, Apr 2010.

\bibitem{Sauls2011PhysRevB.84.214509}
J.~A. Sauls.
\newblock {Surface states, edge currents, and the angular momentum of chiral
  $p$-wave superfluids}.
\newblock {\em Phys. Rev. B}, 84:214509, Dec 2011.

\bibitem{Bouhon2014PhysRevB.90.220511}
Adrien Bouhon and Manfred Sigrist.
\newblock {Current inversion at the edges of a chiral $p$-wave superconductor}.
\newblock {\em Phys. Rev. B}, 90:220511(R), Dec 2014.

\bibitem{higashitani2014spin}
Seiji Higashitani.
\newblock {Spin Current as a Manifestation of Surface Odd-Frequency Pairing in
  Superfluid $^3$He}.
\newblock {\em Journal of the Physical Society of Japan}, 83(7):075002, 2014.

\bibitem{TanakaPhysRevB.72.140503}
Y.~Tanaka, Y.~Asano, A.~A. Golubov, and S.~Kashiwaya.
\newblock {Anomalous features of the proximity effect in triplet
  superconductors}.
\newblock {\em Phys. Rev. B}, 72:140503(R), Oct 2005.

\bibitem{Yokoyama2011PhysRevLett.106.246601}
Takehito Yokoyama, Yukio Tanaka, and Naoto Nagaosa.
\newblock {Anomalous Meissner Effect in a Normal-Metal--Superconductor Junction
  with a Spin-Active Interface}.
\newblock {\em Phys. Rev. Lett.}, 106:246601, Jun 2011.

\bibitem{Higashitani2013PhysRevLett.110.175301}
S.~Higashitani, H.~Takeuchi, S.~Matsuo, Y.~Nagato, and K.~Nagai.
\newblock {Magnetic Response of Odd-Frequency $s$-Wave Cooper Pairs in a
  Superfluid Proximity System}.
\newblock {\em Phys. Rev. Lett.}, 110:175301, Apr 2013.

\bibitem{Asano2015PhysRevB.92.224508}
Yasuhiro Asano and Akihiro Sasaki.
\newblock Odd-frequency cooper pairs in two-band superconductors and their
  magnetic response.
\newblock {\em Phys. Rev. B}, 92:224508, Dec 2015.

\bibitem{kangPhysRevLett.122.095301}
Seji Kang, Sang~Won Seo, Hiromitsu Takeuchi, and Y.~Shin.
\newblock {Observation of Wall-Vortex Composite Defects in a Spinor
  Bose-Einstein Condensate}.
\newblock {\em Phys. Rev. Lett.}, 122:095301, Mar 2019.

\bibitem{TakeuchiPhysRevLett.126.195302}
Hiromitsu Takeuchi.
\newblock {Quantum Elliptic Vortex in a Nematic-Spin Bose-Einstein Condensate}.
\newblock {\em Phys. Rev. Lett.}, 126:195302, May 2021.

\bibitem{Huh2020PhysRevResearch.2.033471}
SeungJung Huh, Kyungtae Kim, Kiryang Kwon, and Jae-yoon Choi.
\newblock {Observation of a strongly ferromagnetic spinor Bose-Einstein
  condensate}.
\newblock {\em Phys. Rev. Res.}, 2:033471, Sep 2020.

\bibitem{Hamner2011PhysRevLett.106.065302}
C.~Hamner, J.~J. Chang, P.~Engels, and M.~A. Hoefer.
\newblock {Generation of Dark-Bright Soliton Trains in Superfluid-Superfluid
  Counterflow}.
\newblock {\em Phys. Rev. Lett.}, 106:065302, Feb 2011.

\bibitem{KimPhysRevLett.119.185302}
Joon~Hyun Kim, Sang~Won Seo, and Y.~Shin.
\newblock {Critical Spin Superflow in a Spinor Bose-Einstein Condensate}.
\newblock {\em Phys. Rev. Lett.}, 119:185302, Oct 2017.

\bibitem{Law2001PhysRevA.63.063612}
C.~K. Law, C.~M. Chan, P.~T. Leung, and M.-C. Chu.
\newblock {Critical velocity in a binary mixture of moving Bose condensates}.
\newblock {\em Phys. Rev. A}, 63:063612, May 2001.

\bibitem{Zhu2015PhysRevA.91.023633}
Qizhong Zhu, Qing-feng Sun, and Biao Wu.
\newblock {Superfluidity of a pure spin current in ultracold Bose gases}.
\newblock {\em Phys. Rev. A}, 91:023633, Feb 2015.

\bibitem{kawaguchi2012spinor}
Yuki Kawaguchi and Masahito Ueda.
\newblock {Spinor Bose--Einstein Condensates}.
\newblock {\em Physics Reports}, 520(5):253--381, 2012.

\bibitem{PhysRevA.86.013613}
Justin Lovegrove, Magnus~O. Borgh, and Janne Ruostekoski.
\newblock {Energetically stable singular vortex cores in an atomic spin-1
  Bose-Einstein condensate}.
\newblock {\em Phys. Rev. A}, 86:013613, Jul 2012.

\bibitem{PhysRevLett.112.075301}
Justin Lovegrove, Magnus~O. Borgh, and Janne Ruostekoski.
\newblock {Energetic Stability of Coreless Vortices in Spin-1 Bose-Einstein
  Condensates with Conserved Magnetization}.
\newblock {\em Phys. Rev. Lett.}, 112:075301, Feb 2014.

\bibitem{PhysRevA.93.033633}
Justin Lovegrove, Magnus~O. Borgh, and Janne Ruostekoski.
\newblock {Stability and internal structure of vortices in spin-1 Bose-Einstein
  condensates with conserved magnetization}.
\newblock {\em Phys. Rev. A}, 93:033633, Mar 2016.

\bibitem{weiss2019controlled}
Lauren~S Weiss, Magnus~O Borgh, Alina Blinova, Tuomas Ollikainen, Mikko
  M{\"o}tt{\"o}nen, Janne Ruostekoski, and David~S Hall.
\newblock {Controlled creation of a singular spinor vortex by circumventing the
  Dirac belt trick}.
\newblock {\em Nat. Commun.}, 10(1):1--8, 2019.

\bibitem{liu2020phase}
I-Kang Liu, Shih-Chuan Gou, and Hiromitsu Takeuchi.
\newblock {Phase diagram of solitons in the polar phase of a spin-1
  Bose-Einstein condensate}.
\newblock {\em Phys. Rev. Res.}, 2:033506, Sep 2020.

\bibitem{underwood2020properties}
Andrew P.~C. Underwood, D.~Baillie, P.~Blair Blakie, and H.~Takeuchi.
\newblock {Properties of a nematic spin vortex in an antiferromagnetic spin-1
  Bose-Einstein condensate}.
\newblock {\em Phys. Rev. A}, 102:023326, Aug 2020.

\bibitem{PhysRevLett.125.170401}
Stefan Lannig, Christian-Marcel Schmied, Maximilian Pr\"ufer, Philipp Kunkel,
  Robin Strohmaier, Helmut Strobel, Thomas Gasenzer, Panayotis~G. Kevrekidis,
  and Markus~K. Oberthaler.
\newblock {Collisions of Three-Component Vector Solitons in Bose-Einstein
  Condensates}.
\newblock {\em Phys. Rev. Lett.}, 125:170401, Oct 2020.

\bibitem{PhysRevLett.125.030402}
X.~Chai, D.~Lao, Kazuya Fujimoto, Ryusuke Hamazaki, Masahito Ueda, and
  C.~Raman.
\newblock {Magnetic Solitons in a Spin-1 Bose-Einstein Condensate}.
\newblock {\em Phys. Rev. Lett.}, 125:030402, Jul 2020.

\bibitem{PhysRevResearch.3.L012003}
Xiao Chai, Di~Lao, Kazuya Fujimoto, and Chandra Raman.
\newblock {Magnetic soliton: From two to three components with SO(3) symmetry}.
\newblock {\em Phys. Rev. Res.}, 3:L012003, Jan 2021.

\bibitem{2021Takeuchi_PhysRevA.104.013316}
Hiromitsu Takeuchi.
\newblock {Phase diagram of vortices in the polar phase of spin-1 Bose-Einstein
  condensates}.
\newblock {\em Phys. Rev. A}, 104:013316, Jul 2021.

\bibitem{Katsimiga_2021}
G~C Katsimiga, S~I Mistakidis, P~Schmelcher, and P~G Kevrekidis.
\newblock {Phase diagram, stability and magnetic properties of nonlinear
  excitations in spinor Bose{\textendash}Einstein condensates}.
\newblock {\em New Journal of Physics}, 23(1):013015, jan 2021.

\bibitem{isoshima2001quantum}
Tomoya Isoshima, Kazushige Machida, and Tetsuo Ohmi.
\newblock {Quantum vortex in a spinor Bose-Einstein condensate}.
\newblock {\em Journal of the Physical Society of Japan}, 70(6):1604--1610,
  2001.

\bibitem{pethick2008bose}
Christopher~J Pethick and Henrik Smith.
\newblock {\em Bose--Einstein Condensation in Dilute Gases}.
\newblock Cambridge University Press, Cambridge, 2008.

\bibitem{Indekeu2015_PhysRevA.91.033615}
Joseph~O. Indekeu, Chang-You Lin, Nguyen Van~Thu, Bert Van~Schaeybroeck, and
  Tran~Huu Phat.
\newblock {Static interfacial properties of Bose-Einstein-condensate mixtures}.
\newblock {\em Phys. Rev. A}, 91:033615, Mar 2015.

\bibitem{Yu2021PhysRevResearch.3.023043}
Xiaoquan Yu and P.~B. Blakie.
\newblock {Dark-soliton-like magnetic domain walls in a two-dimensional
  ferromagnetic superfluid}.
\newblock {\em Phys. Rev. Res.}, 3:023043, Apr 2021.

\bibitem{Yu2021anomalous}
Xiaoquan Yu and PB~Blakie.
\newblock {Anomalous dynamics of magnetic field-driven propagating magnetic
  domain walls}.
\newblock {\em arXiv preprint arXiv:2104.12967}, 2021.

\bibitem{Barankov2002_PhysRevA.66.013612}
R.~A. Barankov.
\newblock {Boundary of two mixed Bose-Einstein condensates}.
\newblock {\em Phys. Rev. A}, 66:013612, Jul 2002.

\bibitem{Schaeybroeck2008_PhysRevA.78.023624}
Bert Van~Schaeybroeck.
\newblock {Interface tension of Bose-Einstein condensates}.
\newblock {\em Phys. Rev. A}, 78:023624, Aug 2008.

\bibitem{Takeuchi2010PhysRevB.81.094517}
Hiromitsu Takeuchi, Naoya Suzuki, Kenichi Kasamatsu, Hiroki Saito, and Makoto
  Tsubota.
\newblock {Quantum Kelvin-Helmholtz instability in phase-separated
  two-component Bose-Einstein condensates}.
\newblock {\em Phys. Rev. B}, 81:094517, Mar 2010.

\bibitem{WKB}
{ This approximation is a bosonic-quasiparticle version of the
  Wentzel-Kramers-Brillouin (WKB) approximation in quantum mechanics, see,
  also, \cite{takeuchi2018doubly}. }.

\bibitem{takeuchi2018doubly}
Hiromitsu Takeuchi, Michikazu Kobayashi, and Kenichi Kasamatsu.
\newblock {Is a doubly quantized vortex dynamically unstable in uniform
  superfluids?}
\newblock {\em Journal of the Physical Society of Japan}, 87(2):023601, 2018.

\bibitem{Skryabin2000PhysRevA.63.013602}
Dmitry~V. Skryabin.
\newblock {Instabilities of vortices in a binary mixture of trapped
  Bose-Einstein condensates: Role of collective excitations with positive and
  negative energies}.
\newblock {\em Phys. Rev. A}, 63:013602, Dec 2000.

\bibitem{Lundh2006PhysRevA.74.063620}
Emil Lundh and Halvor~M. Nilsen.
\newblock {Dynamic stability of a doubly quantized vortex in a
  three-dimensional condensate}.
\newblock {\em Phys. Rev. A}, 74:063620, Dec 2006.

\bibitem{LandauQuantum}
Problem~4 in~Sec. 23~of L.~D.~Landau and E.~Lifshitz.
\newblock {Quantum Mechanics: Non-relativistic Theory, 3rd ed., Course of
  Theoretical Physics Vol. 3}.
\newblock {\em (Pergamon, New York,1999)}.

\bibitem{leggett2006quantum}
Anthony~James Leggett et~al.
\newblock {\em Quantum Liquids: Bose Condensation and Cooper Pairing in
  Condensed-Matter Systems}.
\newblock Oxford University Press, Oxford, 2006.

\bibitem{MacKay1987}
{ R. S. MacKay, in {\it Hamiltonian Dynamical Systems}, ed. R. S. MacKay and J.
  D. Meiss (Adam Hilger, Bristol, U.K., 1987), p. 137. }.

\bibitem{TakeuchiPhysRevA.88.043612}
Hiromitsu Takeuchi and Kenichi Kasamatsu.
\newblock {Nambu-Goldstone modes in segregated Bose-Einstein condensates}.
\newblock {\em Phys. Rev. A}, 88:043612, Oct 2013.

\bibitem{PhysRevA.86.023622}
Shohei Watabe, Yusuke Kato, and Yoji Ohashi.
\newblock {Excitation transport through a domain wall in a Bose-Einstein
  condensate}.
\newblock {\em Phys. Rev. A}, 86:023622, Aug 2012.

\bibitem{2008Kundu}
P.~K. Kundu and I.~M. Cohen.
\newblock {\em Fluid Mechanics, 4th ed.}
\newblock Academic Press, New York, 2008.

\bibitem{tsubota2013quantum}
Makoto Tsubota, Michikazu Kobayashi, and Hiromitsu Takeuchi.
\newblock {Quantum hydrodynamics}.
\newblock {\em Physics Reports}, 522(3):191--238, 2013.

\bibitem{HayashiPhysRevA.87.063628}
Shinsuke Hayashi, Makoto Tsubota, and Hiromitsu Takeuchi.
\newblock {Instability crossover of helical shear flow in segregated
  Bose-Einstein condensates}.
\newblock {\em Phys. Rev. A}, 87:063628, Jun 2013.

\bibitem{Kim2017PhysRevLett.119.047202}
Se~Kwon Kim and Yaroslav Tserkovnyak.
\newblock {Magnetic Domain Walls as Hosts of Spin Superfluids and Generators of
  Skyrmions}.
\newblock {\em Phys. Rev. Lett.}, 119:047202, Jul 2017.

\bibitem{Kokubo2021_PhysRevA.104.023312}
Haruya Kokubo, Kenichi Kasamatsu, and Hiromitsu Takeuchi.
\newblock {Pattern formation of quantum Kelvin-Helmholtz instability in binary
  superfluids}.
\newblock {\em Phys. Rev. A}, 104:023312, Aug 2021.

\bibitem{SM1}
{ See Supplemental Material for the movie files of the animation. }.

\bibitem{Mermin1976_PhysRevLett.36.832.2}
N.~D. Mermin and Tin-Lin Ho.
\newblock {Circulation and Angular Momentum in the $A$ Phase of Superfluid
  Helium-3}.
\newblock {\em Phys. Rev. Lett.}, 36:832--832, Apr 1976.

\bibitem{Skyrme1961_Proc.R.Soc.A}
T.~H.~R. Skyrme.
\newblock {A non-linear field theory}.
\newblock {\em Proc. R. Soc. A}, 260:127--138, 1961.

\bibitem{Parts1995_PhysRevLett.75.3320}
\"U. Parts, J.~M. Karim\"aki, J.~H. Koivuniemi, M.~Krusius, V.~M.~H. Ruutu,
  E.~V. Thuneberg, and G.~E. Volovik.
\newblock {Phase Diagram of Vortices in Superfluid ${}^{3}$
  $He\ensuremath{-}\mathit{A}$}.
\newblock {\em Phys. Rev. Lett.}, 75:3320--3323, Oct 1995.

\bibitem{volovik2003universe}
Grigory~E Volovik.
\newblock {\em The Universe in a Helium Droplet}, volume 117.
\newblock Oxford University Press on Demand, New York, 2003.

\bibitem{girvin1999quantum}
Steven~M Girvin.
\newblock {The quantum Hall effect: Novel excitations and broken symmetries}.
\newblock In S.~Ouvry F.~David A.~Comtet, T.~Jolicoeur, editor, {\em
  Topological Aspects of Low Dimensional Systems}, pages 53--175.
  Springer-Verlag, Berlin and Les Editions de Physique, Les Ulis, 2000.

\bibitem{Moon1995_PhysRevB.51.5138}
K.~Moon, H.~Mori, Kun Yang, S.~M. Girvin, A.~H. MacDonald, L.~Zheng,
  D.~Yoshioka, and Shou-Cheng Zhang.
\newblock {Spontaneous interlayer coherence in double-layer quantum Hall
  systems: Charged vortices and Kosterlitz-Thouless phase transitions}.
\newblock {\em Phys. Rev. B}, 51:5138--5170, Feb 1995.

\bibitem{Kasamatsu2004_PhysRevLett.93.250406}
Kenichi Kasamatsu, Makoto Tsubota, and Masahito Ueda.
\newblock {Vortex Molecules in Coherently Coupled Two-Component Bose-Einstein
  Condensates}.
\newblock {\em Phys. Rev. Lett.}, 93:250406, Dec 2004.

\bibitem{kasamatsu2005vortices}
Kenichi Kasamatsu, Makoto Tsubota, and Masahito Ueda.
\newblock {Vortices in multicomponent Bose--Einstein condensates}.
\newblock {\em International Journal of Modern Physics B}, 19(11):1835--1904,
  2005.

\bibitem{Ezawa2011_PhysRevB.83.100408}
Motohiko Ezawa.
\newblock {Compact merons and skyrmions in thin chiral magnetic films}.
\newblock {\em Phys. Rev. B}, 83:100408(R), Mar 2011.

\bibitem{Kharkov2017_PhysRevLett.119.207201}
Y.~A. Kharkov, O.~P. Sushkov, and M.~Mostovoy.
\newblock {Bound States of Skyrmions and Merons near the Lifshitz Point}.
\newblock {\em Phys. Rev. Lett.}, 119:207201, Nov 2017.

\bibitem{gao2019creation}
Ningbo Gao, S-G Je, M-Y Im, Jun~Woo Choi, Masheng Yang, Qin-ci Li, TY~Wang,
  S~Lee, H-S Han, K-S Lee, et~al.
\newblock Creation and annihilation of topological meron pairs in in-plane
  magnetized films.
\newblock {\em Nat. Commun.}, 10(1):1--9, 2019.

\bibitem{gao2020fractional}
Shang Gao, H~Diego Rosales, Flavia A~G{\'o}mez Albarrac{\'\i}n, Vladimir
  Tsurkan, Guratinder Kaur, Tom Fennell, Paul Steffens, Martin Boehm, Petr
  {\v{C}}erm{\'a}k, Astrid Schneidewind, et~al.
\newblock {Fractional antiferromagnetic skyrmion lattice induced by anisotropic
  couplings}.
\newblock {\em Nature}, 586(7827):37--41, 2020.

\bibitem{jani2021antiferromagnetic}
Hariom Jani, Jheng-Cyuan Lin, Jiahao Chen, Jack Harrison, Francesco
  Maccherozzi, Jonathon Schad, Saurav Prakash, Chang-Beom Eom, Ariando Ariando,
  Thirumalai Venkatesan, et~al.
\newblock {Antiferromagnetic half-skyrmions and bimerons at room temperature}.
\newblock {\em Nature}, 590(7844):74--79, 2021.

\bibitem{lounasmaa1999vortices}
Olli~V Lounasmaa and Erkki Thuneberg.
\newblock {Vortices in rotating superfluid 3He}.
\newblock {\em Proceedings of the National Academy of Sciences},
  96(14):7760--7767, 1999.

\bibitem{seppala1983evidence}
Harri~Kalervo Sepp{\"a}l{\"a} and GE~Volovik.
\newblock {Evidence for nonsingular vorticity in the Helsinki experiments on
  rotating 3 He-A}.
\newblock {\em Journal of low temperature physics}, 51(3):279--290, 1983.

\bibitem{Seppala1984_PhysRevLett.52.1802}
H.~K. Sepp\"al\"a, P.~J. Hakonen, M.~Krusius, T.~Ohmi, M.~M. Salomaa, J.~T.
  Simola, and G.~E. Volovik.
\newblock {Continuous Vortices with Broken Symmetry in Rotating Superfluid
  $^{3}\mathrm{He}$-$A$}.
\newblock {\em Phys. Rev. Lett.}, 52:1802--1805, May 1984.

\bibitem{Volovic1984_PhysRevB.29.6090}
V.~Z. Vulovic, D.~L. Stein, and A.~L. Fetter.
\newblock {NMR of textures in rotating $^{3}\mathrm{He}\ensuremath{-}A$}.
\newblock {\em Phys. Rev. B}, 29:6090--6095, Jun 1984.

\bibitem{Simola1987_PhysRevLett.58.904}
J.~T. Simola, L.~Skrbek, K.~K. Nummila, and J.~S. Korhonen.
\newblock {Two different vortex states in rotating $^{3}\mathit{A}$ observed by
  use of negative ions}.
\newblock {\em Phys. Rev. Lett.}, 58:904--907, Mar 1987.

\bibitem{fetter1987vortex}
Alexander~L Fetter.
\newblock {Vortex structures in rotating 3 He-A}.
\newblock {\em Journal of low temperature physics}, 67(3):145--153, 1987.

\bibitem{Williamson2016PhysRevLett.116.025301}
Lewis~A. Williamson and P.~B. Blakie.
\newblock {Universal Coarsening Dynamics of a Quenched Ferromagnetic Spin-1
  Condensate}.
\newblock {\em Phys. Rev. Lett.}, 116:025301, Jan 2016.

\bibitem{Williamson2016PhysRevA.94.023608}
Lewis~A. Williamson and P.~B. Blakie.
\newblock {Coarsening and thermalization properties of a quenched ferromagnetic
  spin-1 condensate}.
\newblock {\em Phys. Rev. A}, 94:023608, Aug 2016.

\bibitem{Shitara_2017}
Nanako Shitara, Shreya Bir, and P~Blair Blakie.
\newblock {Domain percolation in a quenched ferromagnetic spinor condensate}.
\newblock {\em New Journal of Physics}, 19(9):095003, sep 2017.

\bibitem{Bourges2017_PhysRevA.95.023616}
Andr\'eane Bourges and P.~B. Blakie.
\newblock {Different growth rates for spin and superfluid order in a quenched
  spinor condensate}.
\newblock {\em Phys. Rev. A}, 95:023616, Feb 2017.

\bibitem{takeuchi2016domain}
Hiromitsu Takeuchi.
\newblock {Domain size distribution in segregating binary superfluids}.
\newblock {\em Journal of Low Temperature Physics}, 183(3):169--174, 2016.

\bibitem{TakeuchiPhysRevA.97.013617}
Hiromitsu Takeuchi.
\newblock {Domain-area distribution anomaly in segregating multicomponent
  superfluids}.
\newblock {\em Phys. Rev. A}, 97:013617, Jan 2018.

\end{thebibliography}


\end{document}